\documentclass{jfm}

\usepackage[dvipsnames]{xcolor}

\usepackage{graphicx}
\usepackage{epstopdf, epsfig}
\usepackage{amsmath}
\usepackage[utf8]{inputenc} 

\usepackage{subfigure}% in preamble

\usepackage{pgfplots}
\usepackage{tikz}
\usetikzlibrary{backgrounds,pgfplots.groupplots,external}

\usepackage{lineno}
\shorttitle{Single--to--double-crest wave transition in orbital shaken cylindrical containers}

\shortauthor{A. Bongarzone, M. Guido and F. Gallaire}% M. Farhat

\title{An amplitude equation modeling the single--to--double crest wave transition\\ in orbital shaken cylindrical containers} %Title of paper

\author{Alessandro Bongarzone\aff{1},
  Margherita Guido\aff{2,3} 
 \and François Gallaire\aff{1}
 \corresp{\email{francois.gallaire@epfl.ch}}}

\affiliation{\aff{1}Laboratory of Fluid Mechanics and Instabilities, École Polytechnique Fédérale de Lausanne, Lausanne, CH-1015, Switzerland
\aff{2}Swiss Plasma Center, École Polytechnique Fédérale de Lausanne, Lausanne, CH-1015, Switzerland
\aff{3}Numerical Algorithms and High-Performance Computing, École Polytechnique Fédérale de Lausanne, Lausanne, CH-1015, Switzerland}

\begin{document}

\maketitle
%\tableofcontents

\begin{abstract}

The container motion along a planar circular trajectory at a constant angular velocity, i.e. \textit{orbital shaking}, is of interest in several industrial applications, e.g. for fermentation processes or in cultivation of stem cells, where good mixing and efficient gas exchange are the main targets. Under these external forcing conditions, the free surface typically exhibits a primary steady state motion through a single--crest dynamics, whose wave amplitude, as a function of the external forcing parameters, shows a Duffing-like behaviour. However, previous experiments in lab-scale cylindrical containers have unveiled that, owing to the excitation of super-harmonics, diverse dynamics are observable in certain driving-frequency ranges. Among these super-harmonics, the double-crest dynamics is particularly relevant, as it displays a notably large amplitude response, that is strongly favored by the spatial structure of the external forcing. In the inviscid limit and with regards to circular cylindrical containers, we formalize here a weakly nonlinear analysis via multiple timescale method of the full hydrodynamic sloshing system, leading to an amplitude equation suitable to describe such a super-harmonic dynamics and the resulting single--to--double crest wave transition. The weakly nonlinear prediction is shown to be in fairly good agreement with previous experiments described in the literature. Lastly, we discuss how an analogous amplitude equation can be derived by solving asymptotically for the first super-harmonic of the forced Helmholtz--Duffing equation with small nonlinearities.

%predictive tool for the threshold of sub-harmonic parametric waves and their nonlinear amplitude saturation  

\end{abstract}

%sloshing, amplitude equation
\begin{keywords}
\end{keywords}

%%%%%%%%%%%%%%%%%%%%%%%%%%%%%%%%%%%%%%%%%%%%%%%%%%%%%%%%
%%%%%%%%%%%%%%%%%%%%%%%%%%%%%%%%%%%%%%%%%%%%%%%%%%%%%%%%
%%%%%%%%%%%%%%%%%%%%%%%%%%%%%%%%%%%%%%%%%%%%%%%%%%%%%%%%
%%%%%%%%%%%%%%%%%%%%%%%%%%%%%%%%%%%%%%%%%%%%%%%%%%%%%%%%

\section{Introduction}\label{sec:Intro}

Orbital shaking is a method to gently mix the liquid content of a container by its displacement at fixed container orientation along a circular trajectory and at a constant angular velocity. It is used in biological and chemical industrial applications, notably bacterial and cellular cultures \citep{mcdaniel1969effect,wurm2004production}, as an alternative to stirred tanks, where the liquid agitation results from a rotating impeller or the rotation of magnetic rod. In these cultivation protocols, cells are in suspension in the extracellular liquid medium, which serves as buffer for consumables from which they feed and for their secretions. The motion of the liquid prevents sedimentation and homogenizes the concentration of dissolved oxygen and nutrients and of secreted proteins and carbon dioxide. Thanks to the possible gas exchanges at the free surface, oxygen supply from the container bottom can possibly be circumvented, avoiding the formation of bubbles and thereby the damages that their collapse can exert on cells \citep{handa1989effect,kretzmer1991response,papoutsakis1991fluid}, sparking interest in the development of large-scale, in the hectoliter range, orbital shaken bioreactors \citep{liu2001development,de2004tubespin,muller2007scalable}. It is therefore not a surprise if a significant body of research on the gas exchange and mixing in these devices has emerged over the last two decades \citep{buchs2000power1,buchs2000power2,buchs2001introduction,maier2004advances,muller2005orbital,micheletti2006fluid,zhang2009efficient,tissot2010determination,tan2011measurement,tissot2011efficient,klockner2012advances}.\\
\indent At a more fundamental level, the hydrodynamics of these orbital shaking devices has received recent attention, from both experimental \citep{reclari2014surface,bouvard2017mean,moisy2018counter} and theoretical \citep{reclari2014surface,horstmann2020linear} perspectives, predominantly using linear potential flow models. These models are often complemented with effective viscous damping rates to incorporate the energy dissipation responsible for the phase-shifts between wave and shaker, which was also seen to be sometimes responsible for damping-induced symmetry-breaking linear mechanisms resulting in linear spiral wave patterns  \citep{horstmann2020linear,horstmann2021formation}. Previous studies make mostly use of classical existing theories for general linear sloshing dynamics, reviewed for instance in \cite{ibrahim2009liquid} or \cite{faltinsen2005liquid}.\\
\indent In order to refine the linear potential model and, specifically, to predict the occurrence of the super-harmonic wave dynamics observed experimentally, \cite{reclari2013hydrodynamics} and \cite{reclari2014surface} proposed an inviscid weakly nonlinear analysis based on a second order straightforward asymptotic expansion procedure, which was shown to be capable of capturing the observed resonance frequencies and of characterizing different multiple--crest wave patterns. Among these patterns, the super-harmonic double--crest wave dynamics is particularly relevant, as it appears to be the most stable and the one which displays the largest amplitude response. However, their analysis, as typical of straightforward asymptotic expansions, suffers from secular terms \citep{castaing2005hydrodynamics,Nayfeh2008} and, therefore, it still fails in describing the correct nonlinear behaviour close to both harmonic and super-harmonic resonances.\\ 
\indent This limitation was partially overcome by \cite{timokha2017} and \cite{raynovskyy2018steady}, who have applied the Narimanov–Moiseev multimodal sloshing theory \citep{narimanov1957movement,moiseev1958theory,dodge1965liquid,faltinsen1974nonlinear,narimanov1977,lukovsky1990} describing the nonlinear wave dynamics near the first harmonic resonance, which was proven to be of the \textit{hard}-spring type, in qualitative agreement with the observations of \cite{reclari2013hydrodynamics} and \cite{reclari2014surface}. Nevertheless, they were not able to quantitatively compare their predictions with the latter experimental measurements because these were done for $H=h/R\leq1.2$, e.g. $H=1$ and $1.04$ (with $h$ the fluid depth and $R$ the container radius), where, as stated by the authors themselves, the adopted nonlinear modal equations are not applicable due to the secondary resonance phenomena \citep{faltinsen2016resonant}.\\
\indent In the spirit of the aforementioned multimodal theory but with regards to square-base basins, the resonant amplification of higher order modes was investigated by \cite{faltinsen2005resonant}, who formalized a so-called adaptive asymptotic modal approach capable to improve the agreements with earlier experiments. Yet, to the knowledge of the authors, the adaptive modal approach was never extended to orbital shaken circular cylindrical containers.\\
\indent For these reasons, it appears that a quantitatively accurate model for the prediction of the diverse wave dynamics observed during the thorough experimental campaign carried out by \cite{reclari2013hydrodynamics} and \cite{reclari2014surface} has not been provided yet.\\
\indent The present work is precisely dedicated to the development of a weakly nonlinear analysis based on the multiple timescale method, which will be seen to successfully capture nonlinear effects for the main additive harmonic resonances as well as the more subtle additive and multiplicative resonance governing the super-harmonic double--crest saturation. Amplitude equations are rigorously derived in an inviscid framework, which once amended with an \textit{ad-hoc} damping term as only tuning parameter, well match the experimental findings of \cite{reclari2013hydrodynamics} and \cite{reclari2014surface}. Lastly, the obtained amplitude equations for harmonic single--crest and super-harmonic double--crest waves are found to be compatible with the two well known one-degree-of-freedom (1dof) systems, the Duffing and the Helmholtz-Duffing oscillators, respectively.\\
\indent The manuscript is organized as follows. The flow configuration and governing equations are introduced in \S\ref{sec:Sec1}. \S\ref{sec:Sec2} is dedicated to briefly summarize the salient points of the asymptotic model proposed by \cite{reclari2014surface}, whose limitations motivated the present work. After tackling the more common case of harmonic single--crest wave in \S\ref{subsec:Sec3sub1}, the weakly nonlinear amplitude equation governing the super-harmonic double--crest wave dynamics is derived in \S\ref{subsec:Sec3sub2}. Final comments and conclusions are outlined in \S\ref{sec:Conc}.\\

%\cite{hutton1963investigation}
%\cite{miles1976nonlinear}
%\cite{faltinsen2000multidimensional,faltinsen2003resonant,faltinsen2006resonant,faltinsen2017resonant,faltinsen2020resonant}

\section{Flow configuration and governing equations: potential model}\label{sec:Sec1}

 \begin{figure}
\centering
\includegraphics[scale=0.4]{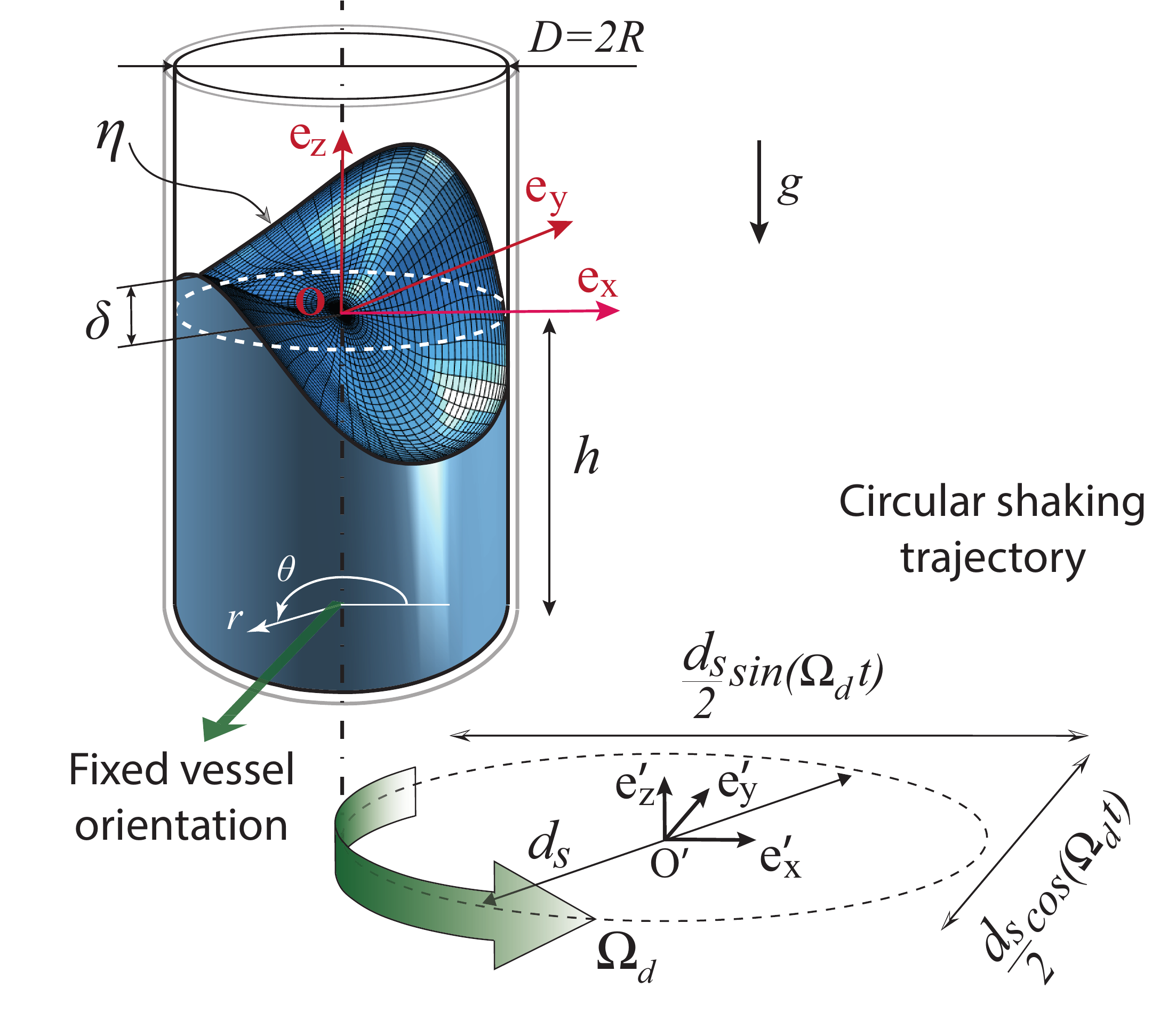}% Here is how to import EPS art
\caption{Sketch of a cylindrical container of diameter $D=2R$ and filled to a depth $h$. The gravity acceleration is denoted by $g$. $O'\mathbf{e}'_x\mathbf{e}'_y\mathbf{e}'_z$ is the Cartesian inertial reference frame, while $O\mathbf{e}_x\mathbf{e}_y\mathbf{e}_z$ is the Cartesian reference frame moving with the container. The origin of the moving cylindrical reference frame $\left(r,\theta,z\right)$ is placed at the container revolution axis and, specifically, at the unperturbed liquid height, $z=0$. The perturbed free surface and contact line elevation are denoted by $\eta$ and $\delta$, respectively. $d_s$ is the diameter of the circular shaking trajectory, characterized by a driving angular frequency $\Omega_d$.}
\label{fig:Fig0} 
\end{figure}

We consider a cylindrical container of diameter $D=2R$ filled to a depth $h$ with a liquid of density $\rho$. The air--liquid surface tension is denoted by $\gamma$. The orbital (circular) shaking motion (see sketch in figure~\ref{fig:Fig0}) can be represented as the combination of two sinusoidal translations with a $\pi/2$ phase shift, thus leading to the following equations of motion for the container axis intersection with the $z=0$ plane, parametrized in cylindrical coordinates ($r$, $\theta$)
\begin{equation}
\label{eq:EqMotWall}
\dot{\mathbf{X}}_0 =
  \begin{cases}
    -\frac{d_s}{2}\Omega_d\sin{\left(\Omega_d t-\theta\right)}\,\mathbf{e}_r\\
    \frac{d_s}{2}\Omega_d\cos{\left(\Omega_d t-\theta\right)}\,\mathbf{e}_{\theta}
  \end{cases}.
\end{equation}
In the classical potential flow limit, i.e. the flow is assumed to be inviscid, irrotational and incompressible, the motion is described in terms of free surface deformation, $\eta$, and a potential velocity field, $\Phi_{tot}$, which is typically separated into a container, $\Phi_c$, and a fluid component, $\Phi$. Hence, the liquid motion within the moving container is governed by the Laplace equation,
\begin{equation}
\label{eq:GovEq_Lap}
\Delta\Phi=\frac{1}{r}\frac{\partial\Phi}{\partial r}+\frac{\partial^2\Phi}{\partial r^2}+\frac{1}{r^2}\frac{\partial^2\Phi}{\partial \theta^2}+\frac{\partial^2\Phi}{\partial z^2}=0,
\end{equation}
subjected to the homogeneous no-penetration condition, $\nabla\Phi\cdot\mathbf{n}=\mathbf{0}$, at the solid sidewall and bottom, and by the dynamic and kinematic free surface boundary conditions at $z=\eta$ (see \cite{ibrahim2009liquid}),
\begin{subequations}
\begin{equation}
\label{eq:GovEq_Dyn}
\frac{\partial\Phi}{\partial t}+\frac{1}{2}\nabla\Phi\cdot\nabla\Phi+\eta-\frac{\kappa\left(\eta\right)}{Bo}=rf\cos{\left(\Omega t-\theta\right)},
\end{equation}
\begin{equation}
\label{eq:GovEq_Kin}
\frac{\partial \eta}{\partial t}+\frac{\partial\Phi}{\partial r}\frac{\partial \eta}{\partial r}+\frac{1}{r^2}\frac{\partial\Phi}{\partial \theta}\frac{\partial \eta}{\partial \theta}-\frac{\partial\Phi}{\partial z}=0,
\end{equation}
\end{subequations}
which have been made non-dimensional by using the container's characteristic length $R$, the characteristic velocity $\sqrt{gR}$ and the time scale $\sqrt{R/g}$. In~\eqref{eq:GovEq_Dyn}, $\kappa\left(\eta\right)$ denotes the fully nonlinear curvature, while $Bo=\rho gR^2/\gamma$ is the Bond number. The non-dimensional driving amplitude and angular frequency read $f=d_s\Omega_d^2/\left(2g\right)$ and, $\Omega=\Omega_d/\sqrt{g/R}$, respectively. When surface tension is accounted for, an additional contact line boundary condition is required at $z=\eta$ and $r=1$, typically written as $\partial\eta/\partial r=\cot{\vartheta}$, where $\vartheta$ is the macroscopic contact angle. Under the classic free--end edge contact line assumption with $\vartheta=\pi/2$ adopted here, the latter dynamic equation simply reduces to $\partial\eta/\partial r=0$. This means that the free surface at rest is flat and that a $\pi/2$ static contact angle is maintained when the contact line elevation changes dynamically.

\section{Linear solution and second-order straightforward asymptotic expansion}\label{sec:Sec2}

In order to enlighten the limitations of the expansion procedure developed by \cite{reclari2014surface}, which motivates the formalization of the new theoretical framework proposed in the present paper, we briefly recall the salient points. 
Let us consider the following asymptotic expansion for the flow quantities,
\begin{subequations}
\begin{equation}
\label{eq:straight_asymp_exp_Phi}
\Phi=\Phi_0+\epsilon\Phi_1+\epsilon^2\Phi_2+\text{O}\left(\epsilon^3\right),
\end{equation}
\begin{equation}
\label{eq:straight_asymp_exp_xi}
\eta=\eta_0+\epsilon\eta_1+\epsilon^2\eta_2+\text{O}\left(\epsilon^3\right),
\end{equation}
\end{subequations}
together with the further assumption of small driving forcing amplitudes of order $\text{O}\left(\epsilon\right)$, i.e. $f=\epsilon F$, with $\epsilon$ a small parameter $\epsilon\ll 1$. Solution $\mathbf{q}_0=\left(\Phi_0,\eta_0\right)$ represents the rest state, which has a potential velocity field null everywhere, $\Phi_0=0$, and a flat interface, $\eta_0=0$, as the contact angle is here assumed to be $\vartheta=\pi/2$. Substituting the expansions above in equations~\eqref{eq:GovEq_Lap}-\eqref{eq:GovEq_Kin}, a series of system at the various order in $\epsilon$ is obtained. At leading order, equations~\eqref{eq:GovEq_Lap}-\eqref{eq:GovEq_Kin} reduce to a forced linear system, whose matrix compact form reads,
\begin{equation}
\label{eq:LinMatForm}
\left(\partial_t\mathcal{B}-\mathcal{A}\right)\mathbf{q}_1=\boldsymbol{\mathcal{F}}_1,
\end{equation}
with $\mathbf{q}_1=\left\{\Phi_1,\eta_1\right\}^T$, $\boldsymbol{\mathcal{F}}_1=F\left\{0,r/2\right\}^Te^{\text{i}\left(\Omega t-\theta\right)}+c.c.=F\boldsymbol{\hat{\mathcal{F}}}_1^Fe^{\text{i}\left(\Omega t-\theta\right)}+c.c.$ and
\begin{equation}
    \label{eq:LinMatForm_Expr}
    \mathcal{B} = 
 \begin{pmatrix}
  0 & 0 \\
  I_{\eta} & 0\\
 \end{pmatrix},
    \mathcal{A} = 
 \begin{pmatrix}
  \Delta & 0 \\
  0 & -I_{\eta}+\frac{1}{Bo}\frac{\partial\kappa\left(\eta\right)}{\partial\eta}\\
 \end{pmatrix},
 \end{equation}
where $c.c.$ stands for complex conjugate, $\partial\kappa\left(\eta\right)/\partial\eta$ represents the first order variation of the curvature associated with the small perturbation $\epsilon\eta_1$ and $I_{\eta}$ is the identity matrix associated with the interface $\eta$. Note that the kinematic condition does not explicitly appear in~\eqref{eq:LinMatForm_Expr}, but it is enforced as a boundary condition at the interface \citep{Viola2018a}. %bongarzone2021faraday 
In the limit of zero external forcing, i.e. $F=0$, system~\eqref{eq:LinMatForm} is a linear homogeneous problem which, by seeking for solutions having the following normal form 
\begin{equation}
\label{eq:GenEigProb0}
\hat{\mathbf{q}}^{mn}\left(r,z\right)e^{\text{i}\left(\omega_{mn} t-m\theta\right)}+c.c.,
\end{equation}
reduces to the classic generalized eigenvalue problem for inviscid capillary--gravity waves
\begin{equation}
\label{eq:GenEigProb1}
\left(\text{i}\omega_{mn}\mathcal{B}-\mathcal{A}_m\right)\hat{\mathbf{q}}^{mn}=\mathbf{0},
\end{equation}
where indices $\left(m,n\right)$ represent the number of nodal circles and nodal diameters, respectively, with $m$ also commonly known as azimuthal wavenumber. Owing to the normal mode expansion, we notice that the operator $\mathcal{A}_m$ is complex, since $\theta$ derivatives produce $-\text{i}m$ terms. An exact analytical solution to equation~\eqref{eq:GenEigProb1} can be readily obtained via a Bessel-Fourier-series representation leading to the well-known dispersion relation \citep{Lamb32}
\begin{equation}
\label{eq:DispRel}
\omega_{mn}^2=\left(k_{mn}+k_{mn}^3/Bo\right)\tanh{\left(k_{mn}H\right)},
\end{equation}
with $H=h/R$ and where the wavenumbers $k_{mn}$ is given by the \textit{n}th-root of the first derivative of the \textit{m}th-order Bessel function of the first kind satisfying $J'_{m}\left(k_{mn}\right)=0$.\\
\indent Despite the existence of this analytical solution, in this work we opt for a numerical scheme based on a discretization technique, where linear operators $\mathcal{B}$ and $\mathcal{A}_m$ are discretized in space by means of a Chebyshev pseudospectral collocation method with a two-dimensional mapping implemented in Matlab, which is analogous to that described by \cite{Viola2018a}. This numerical technique will enable us to avoid straightforward, but cumbersome calculations, otherwise required in the development of the rest of this work and, particularly, of section \S\ref{subsec:Sec3sub2}. One must note that when~\eqref{eq:GenEigProb1} is solved numerically as in the present case, additional boundary conditions need to be made explicit in order to regularize the problem on the revolution axis ($r=0$), i.e.
\begin{subequations}
\begin{equation}
\label{eq:BCm0}
m=0:\ \ \ \ \ \frac{\partial\hat{\eta}^{mn}}{\partial r}=\frac{\partial\hat{\Phi}^{mn}}{\partial r}=0,
\end{equation}
\begin{equation}
|m|\ge1:\ \ \ \ \ \ \ \ \ \ \hat{\eta}^{mn}=\hat{\Phi}^{mn}=0.
\label{eq:BCm1}
\end{equation}
\end{subequations}
It is also useful to note that owing to the symmetries of the problem, system~\eqref{eq:GenEigProb1} is invariant under the transformation
\begin{equation}
\left(\hat{\mathbf{q}}^{mn},+m,\text{i}\omega_{mn}\right)\longrightarrow\left(\hat{\mathbf{q}}^{mn},-m,\text{i}\omega_{mn}\right).
\end{equation}
\indent Convergence of the numerical solution was checked by computing the first 16 modes ($m=0,2,3,4$ with $n=1,2,3,4$), whose corresponding natural frequency values, $\omega_{mn}$, matched the analytical ones given by~\eqref{eq:DispRel} up to the fourth digit for a computational grid $N_r=N_z=60$, with $N_r$ and $N_z$ the number of radial and axial grid points, respectively.\\
\indent Let us now reintroduce the forcing term on the r.h.s. of equation~\eqref{eq:LinMatForm}. In contradistinction with the cases of unidirectional forcing \citep{miles1984internally,miles1984resonantly}, for circular orbits, given the azimuthal periodicity of the associated forcing, the shaking at linear order is expected to excite non-axisymmetric modes only and, specifically, those with $m=1$. Therefore, the linear response to the external forcing can be sought as
\begin{equation}
\label{eq:WNL_Forc_eps1}
\mathbf{q}_{1}=F\hat{\mathbf{q}}_1^Fe^{\text{i}\left(\Omega t-\theta\right)}+c.c.,
\end{equation}
with $\hat{\mathbf{q}}_1^F$ being the solution of the following forced problem
\begin{equation}
\label{eq:WNL_Forc_sys_eps1}
\left(\text{i}\Omega\mathcal{B}-\mathcal{A}_{1}\right)\hat{\mathbf{q}}_1^F=\boldsymbol{\hat{\mathcal{F}}}_1^F.
\end{equation}
The response structure $\hat{\mathbf{q}}_1^F$ is here computed numerically, but, in practice, it is formally equivalent to that obtained analytically by \cite{reclari2014surface} by projecting the forcing term $\boldsymbol{\hat{\mathcal{F}}}_1$ onto the basis formed by the first order Bessel functions of the first kind, except that surface tension is retained here because of the finite Bond number. 
\begin{figure}
\centering
\includegraphics[width=\textwidth]{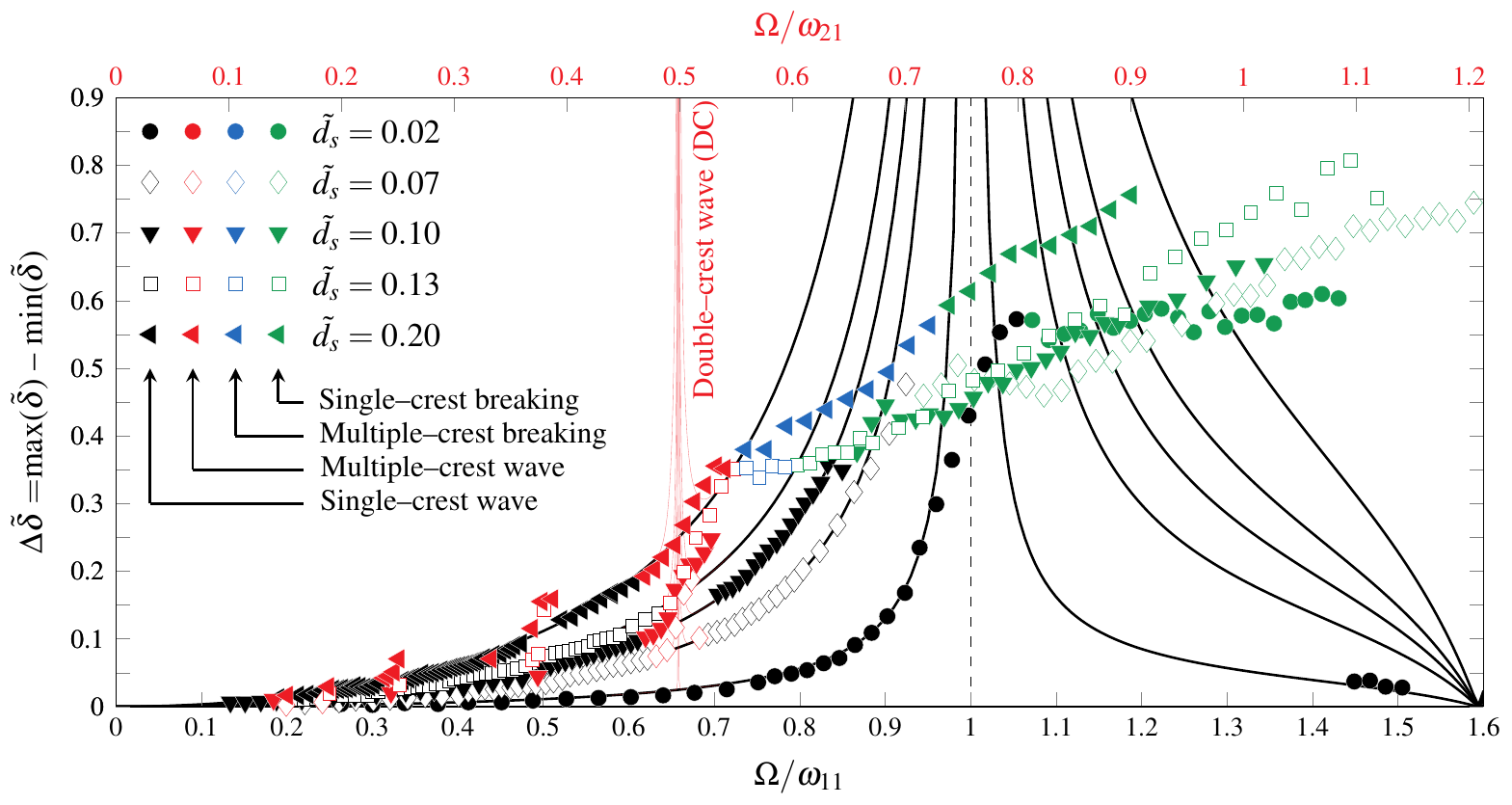}
\caption{Markers correspond to the experimentally measured maximum crest-to-trough contact line amplitude (non-dimensional), with $\tilde{\delta}=\delta R/D=\delta/2$, reported by \cite{reclari2014surface} for two container diameters, $D=0.144\,\text{m}$ and $D=0.287\,\text{m}$, a non-dimensional depth $\tilde{H}=h/D=0.52$ and five values of $\tilde{d}_s=d_s/D$, as a function of the non-dimensional shaking frequency $\Omega$ normalized by the natural frequency of the first non-axisymmetric mode, $\omega_{11}=1.3286$ ($m=1$) on the bottom-x-axis and by that of first non-axisymmetric mode with $m=2$, $\omega_{21}=1.7475$, on the top-x-axis (the frequency values correspond to $D=0.287\,\text{m}$). Colors denote different wave conditions. Black solid lines: linear potential model solution, from~\eqref{eq:WNL_Forc_eps1}, computed by solving numerically equation~\eqref{eq:WNL_Forc_sys_eps1}. Red solid lines: weakly nonlinear solution close to the $\Omega\approx \omega_{21}/2$, obtained by computing~\eqref{eq:WNLS_Forc_Tot}. Note that in order to ease the comparison with experiments, the non-dimensional $\delta$ was rescaled by a factor $R/D=1/2$, as the container diameter $D$, rather than the container radius $R$, was used by \cite{reclari2014surface} to make the equations non-dimensional.}
\label{fig:Fig1} 
\end{figure}
Noting that $\epsilon F=f=d_s\Omega^2/\left(2g\right)$, in figure~\ref{fig:Fig1} the linear solution $\epsilon\mathbf{q}_1^F$ from~\eqref{eq:WNL_Forc_eps1} is shown (black solid lines) and compared with experimental measurements reported by \cite{reclari2014surface} in terms of maximum non-dimensional crest-to-trough contact line amplitudes, $\tilde{\delta}=\delta R/D$, with $\delta\left(\theta,t\right)=\eta\left(r=1,\theta,t\right)$. Measurements for different values of the non-dimensional shaking diameters, $\tilde{d}_s=d_s/D$, are shown. Blue and green markers in figure~\ref{fig:Fig1} correspond to highly nonlinear scenarios manifesting a free surface breaking, which will be therefore ignored thereafter. As discussed by \cite{reclari2014surface} and reproduced here, the linear solution describes well the single--crest (SC) wave dynamics for driving frequencies far enough from harmonic resonances and, particularly, for small $\tilde{d}_s$. However, as typical of undamped forced oscillators, the amplitude of the inviscid linear response to the external forcing is proportional to $\propto 1/\left(\omega_{1n}^2-\Omega^2\right)$ and therefore it naturally diverges close to $\omega_{1n}$, thus failing in predicting the close-to-resonance behaviour, e.g. for $\tilde{d}_s=0.02$ at $\Omega\approx\omega_{11}$. Introduction of viscous dissipation would regularize the divergent behaviour at $\Omega=\omega_{11}$, however, in absence of any nonlinear restoring term, the \textit{hardening} nonlinearity displayed in figure~\ref{fig:Fig1} cannot be retrieved.\\
\indent Furthermore, in experiments multiple--crested waves were observed at fractions of the natural frequencies (red markers in figure~\ref{fig:Fig1}), i.e. the system responses with a frequency which is \textit{n}-times (with \textit{n} positive integer) that of the external forcing. Here we refer to such conditions as super-harmonic dynamics (note that the terminology sub--harmonic was used by \cite{reclari2014surface} instead). Among these super-harmonics, the double--crest (DC) wave dynamics, occurring at a driving frequency $\Omega\approx\omega_{21}/2$, was seen to be the most relevant (see figure~\ref{fig:Fig1}), i.e. the most stable and the one displaying the largest deviation from the linear approximation. This specific multiple--crest dynamics, which is intrinsically nonlinear, is indeed favored by the azimuthal symmetry of the external forcing. \cite{reclari2014surface} tentatively described such a double--crest dynamics by pursuing the asymptotic analysis up to the second order in $\epsilon$, as in equations~\eqref{eq:straight_asymp_exp_Phi} and \eqref{eq:straight_asymp_exp_xi}, in order to account for second order system weak nonlinearities.\\
\indent At the second order in $\epsilon$, one obtains the following forced linear system,
\begin{equation}
\label{eq:WNLSMatForm}
\left(\partial_t\mathcal{B}-\mathcal{A}\right)\mathbf{q}_2=\boldsymbol{\mathcal{F}}_2=F^2\left(\boldsymbol{\hat{\mathcal{F}}}_2^{FF}e^{\text{i}\left(2\Omega t-2\theta\right)}+c.c.\right)+F^2\boldsymbol{\hat{\mathcal{F}}}_2^{F\overline{F}},
\end{equation}
where $\boldsymbol{\mathcal{F}}_2$ gathers a series of terms produced by the first order solution through the second order system nonlinearities. For the sake of brevity, the explicit expression of these forcing terms is here omitted (see \cite{ibrahim2009liquid} and \cite{reclari2013hydrodynamics}, among others, for a full derivation up to the second and third order). The bar denotes the complex conjugate. Also note that amplitude $F$ is actually a real quantity and in the following the superscript $^{\overline{F}}$ will be used only to indicate forcing terms produced by the combination of the direct and complex conjugate contributions of the first order response to the external forcing. The r.h.s. of equation~\eqref{eq:WNLSMatForm} clearly shows how second order terms naturally induce a super-harmonic response, whose spatial periodicity is $m=2$, hence precisely corresponding to the double--crest dynamics experimentally observed. The second forcing term on the r.h.s. of~\eqref{eq:WNLSMatForm} has $\omega=0$ and $m=0$, i.e. it is steady and axisymmetric. It originates in the leading order contribution in the time and azimuthal averaged flow, the so-called mean flow. Equation~\eqref{eq:WNLSMatForm} was solved analytically by \cite{reclari2014surface} by retaining for convenience only two modes, namely those with $\left(m,n\right)=\left(2,1\right)$ and $\left(0,1\right)$, expected to be the relevant ones. The numerical scheme employed in this work allows us to effortlessly account for all the $\left(2,n\right)$ and $\left(0,n\right)$ modes simultaneously, as their contribution will be directly encompassed in the spatial function $\hat{\mathbf{q}}_2^{FF}$ and $\hat{\mathbf{q}}_2^{F\overline{F}}$, appearing in the second order solution,
\begin{equation}
\label{eq:WNLS_Forc_eps2}
\mathbf{q}_{2}=\left(F^2\hat{\mathbf{q}}_2^{FF}e^{\text{i}\left(2\Omega t-2\theta\right)}+c.c.\right)+F^2\hat{\mathbf{q}}_2^{F\overline{F}},
\end{equation}
whose contributions are computed by solving the following systems
\begin{equation}
\label{eq:WNLS_Forc_eps2_sys1}
\left(\text{i}2\Omega\mathcal{B}-\mathcal{A}_{2}\right)\hat{\mathbf{q}}_2^{FF}=\hat{\boldsymbol{\mathcal{F}}}_2^{FF},\ \ \ \ \ \ -\mathcal{A}_0\hat{\mathbf{q}}_2^{F\overline{F}}=\hat{\boldsymbol{\mathcal{F}}}_2^{F\overline{F}}
\end{equation}
The total flow field, obtained through the asymptotic model is then given by the sum of the first and second order solutions, 
\begin{equation}
\label{eq:WNLS_Forc_Tot}
\mathbf{q}=\left(f\hat{\mathbf{q}}_1^Fe^{\text{i}\left(\Omega t-\theta\right)}+f^2\hat{\mathbf{q}}_2^{FF}e^{\text{i}\left(2\Omega t-2\theta\right)}+c.c.\right)+f^2\hat{\mathbf{q}}_2^{F\overline{F}},
\end{equation}
where, in order to eliminate the implicit small parameter $\epsilon$, the amplitudes $\epsilon F$ and $\epsilon^2 F^2$ are recast in terms of the physical amplitudes, $f$ and $f^2$, respectively. The resulting prediction of the maximum crest-to-trough contact line amplitude, $\delta\left(\theta,t\right)=\eta\left(r=1,\theta,t\right)$ is shown in figure~\ref{fig:Fig1} for driving frequencies close to $\Omega/\omega_{21}\approx 0.5$ (see top-x-axis) as red solid lines. Although this straightforward asymptotic expansion detects the emergence of the super-harmonic double--crest wave in that frequency window, it completely fails in capturing the correct nonlinear wave amplitude saturation displaying a \textit{hardening} behaviour clearly visible in figure~\ref{fig:Fig1}. Once again, the amplitude of the inviscid second harmonic response is proportional to $\propto 1/\left(\omega_{2n}^2-4\Omega^2\right)$ and the total solution tends to diverge close to the double--crest super-harmonic at $\omega_{21}/2$.\\
\indent Such a close-to-resonance divergent behaviour is actually expected (particularly in absence of any form of dissipation) when performing straightforward asymptotic expansions, as they typically suffer from secular (or resonating) terms that must be properly treated (see \cite{castaing2005hydrodynamics} and \cite{Nayfeh2008} among many other references).

\section{Weakly nonlinear analysis via multiple timescale method}\label{sec:Sec3}

In order to overcome the aforementioned limitations of the straightforward asymptotic expansion procedure and thus to attempt to bridge the gap between theoretical predictions and experimental observations, we conduct in this section a weakly nonlinear analysis (WNL) based on the multiple timescale method. With the aim to derive a weakly nonlinear amplitude equation governing the double--crest dynamics (DC), we first tackle the simpler problem of single--crest waves (SC). In both cases we look for a third order asymptotic solution of the system
\begin{equation}
\label{eq:WNL_exp}
\mathbf{q}=\left\{\Phi,\eta\right\}^T=\epsilon\mathbf{q}_1+\epsilon^2\mathbf{q}_2+\epsilon^3\mathbf{q}_3+\text{O}\left(\epsilon^4\right),
\end{equation}
where the zero order solution, $\mathbf{q}_0=\mathbf{0}$, is omitted. 

\subsection{Single--crest dynamics (SC)}\label{subsec:Sec3sub1}

\begin{figure}
\centering
\includegraphics[width=0.8\textwidth]{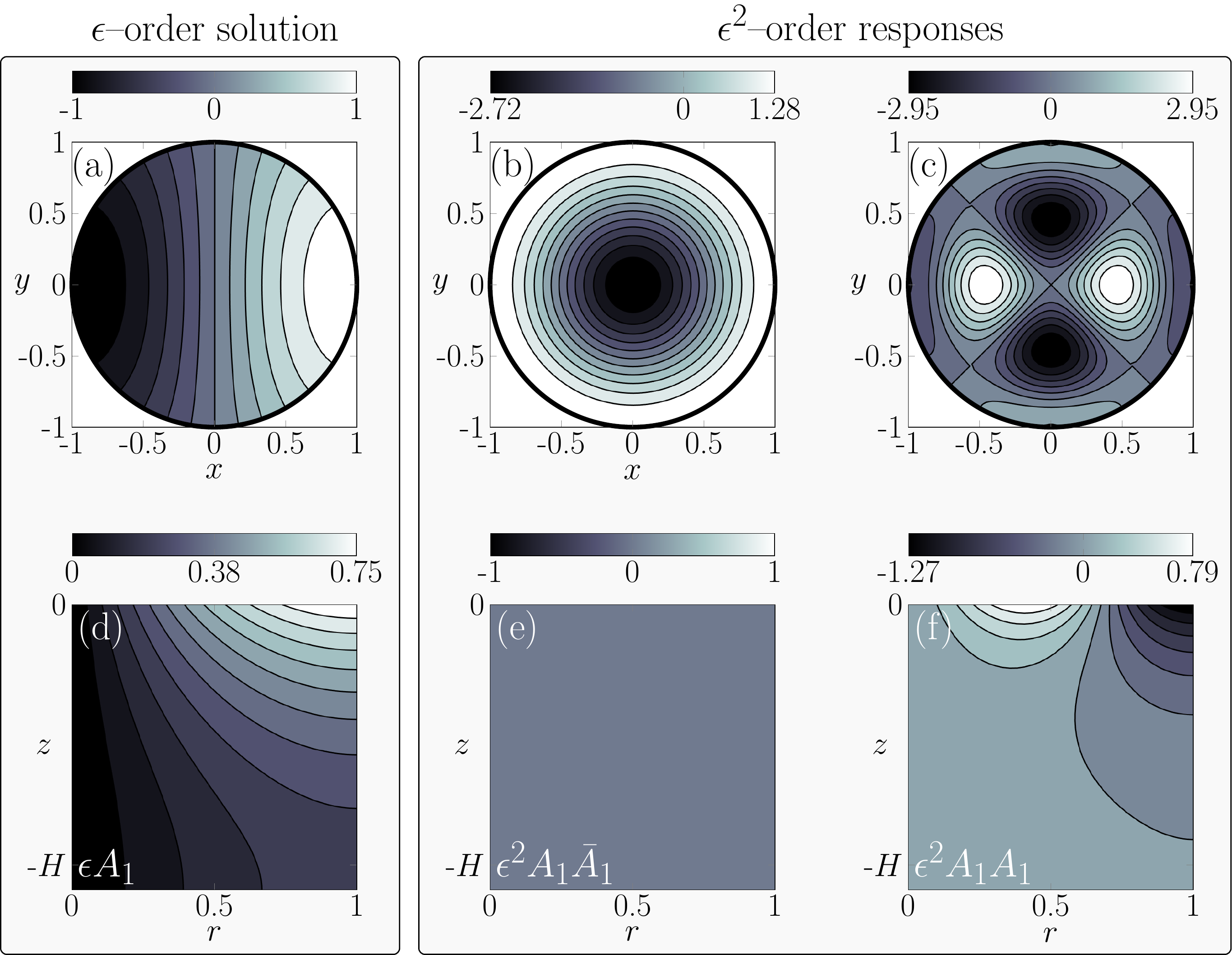}
\caption{(a), (b) and (c): real part of the first, $\hat{\eta}_1^{A_1}$, and second order, $\hat{\eta}_2^{A_1A_1}$ and $\hat{\eta}_2^{A_1\overline{A_1}}$, free surface deformations computed for $\omega_{1n}=\omega_{11}$. (d), (e) and (f): imaginary part of the associated first order, $\hat{\Phi}_1^{A_1}$, and second order, $\hat{\Phi}_2^{A_1A_1}$ and $\hat{\Phi}_2^{A_1\overline{A_1}}$, potential velocity field. Each response is denoted by its amplitude dependence, $\epsilon A_1$, $\epsilon^2 A_1\overline{A}_1$ and $\epsilon^2A_1A_1$. The first order eigenmode is normalized with the amplitude and phase of the contact line (at $r=1$), such that the free surface, $\hat{\eta}_1^{A_1}$ is purely real, whereas $\hat{\Phi}_1^{A_1}$ is purely imaginary. Note that the second order mean flow constantly induces an upside down \textit{bell}-like axisymmetric interface deformation pushing the free surface downward at the center of the moving container. Calculations are performed for the case of figure~\ref{fig:Fig1}, i.e. pure water with $\rho=1000\,\text{m}/\text{m}^3$, $\gamma=0.072\,\text{N}/\text{m}$, $D=0.287\,\text{m}$ and $\tilde{H}=h/D=0.52$, for which $Bo=2\,802.8$ and $\omega_{11}=1.3286$.}
\label{fig:Fig11} 
\end{figure}

In \S\ref{sec:Sec2} the forcing amplitude $f$ was assumed of order $\epsilon$, thus leading to a linear first order problem directly forced by the external shaking, which produces a divergent response close to harmonic resonances. With regards to single--crest waves and specifically to the harmonic response at a driving frequency close to that of one of the non-axisymmetric modes, $\omega_{1n}$, we assume here a small forcing amplitude of order $\epsilon^3$. This assumption is justified by the fact that close-to-resonance, $\Omega\approx\omega_{1n}$, and in absence of dissipation, even a small forcing will induce a large system response. The following analysis is therefore expected to hold for $\Omega=\omega_{1n}+\lambda$, where $\lambda$ is a small detuning parameter, here assumed of order $\epsilon^2$. Lastly, in the spirit of the multiple scale technique, we introduce the slow time scale $T_2 = \epsilon^2t$, with $t$ being the fast time scale at which the free surface oscillates with angular frequency $\approx\omega_{1n}$. Hence, the following scalings are assumed:
\begin{equation}
\label{eq:scaling_SC}
f=\epsilon^3 F,\ \ \ \ \ \lambda=\epsilon^2\Lambda,\ \ \ \ \ T_2=\epsilon^2 t,
\end{equation}
We note that the forcing amplitude could be assumed of order $\epsilon^2$ (as the other parameters), however this complicates unnecessarily the second order problem without modifying the final amplitude equation.\\
\indent Although the asymptotic expansion is here pursued up to the third order in $\epsilon$, the procedure of the weakly nonlinear analysis is essentially equivalent to that of the straightforward asymptotic analysis discussed in \S\ref{sec:Sec2}. The major difference lies in the solution form of the leading order problem that is now a homogenous problem, as in equation~\eqref{eq:GenEigProb1}. Given the azimuthal periodicity of the external forcing, among all possible natural eigenmodes we assume a leading order solution as
\begin{equation}
\label{eq:SC_eps1}
\mathbf{q}_1=A_1\left(T_2\right)\hat{\mathbf{q}}_{1}^{A_1}e^{\text{i}\left(\omega_{1n} t-\theta\right)}+c.c.,
\end{equation}
where $\hat{\mathbf{q}}_1^{A_1}$ is the eigenmode (computed by solving~\eqref{eq:GenEigProb1}) associated with $\left(m,n\right)=\left(1,n\right)$ and $\omega_{1n}$ is the corresponding natural frequency (solution of~\eqref{eq:DispRel}).\\
\indent  The complex amplitude $A_1$, function of the slow time scale $T_2$ and still unknown at this stage of the expansion, describes the slow time amplitude modulation of the oscillating wave $\hat{\mathbf{q}}_1^{A_1}$ and introduces a new arbitrariness in the problem, which must be fixed at a higher order. Eigen-surface, $\hat{\eta}_1^{A_1}$, and eigen-potential field, $\hat{\Phi}_1^{A_1}$, computed for $\omega_{1n}=\omega_{11}$, are shown in figure~\ref{fig:Fig11}(a) and (b), respectively. \\
\indent By pursuing the expansion to the second order, a linear system forced by the first order solution and analogous to that of equation~\eqref{eq:WNLSMatForm} is obtained (see \cite{reclari2013hydrodynamics} for the full expansion of the original nonlinear governing equation up to the second order). Nevertheless, the forcing terms on the r.h.s. are here proportional to $A_1^2$ (super- or second-harmonic) and to $A_1\overline{A}_1$ (mean flow). Thus, we seek for a second order solution of the form
\begin{equation}
\label{eq:SC_eps2}
\mathbf{q}_2=A_1\overline{A}_1\hat{\mathbf{q}}_2^{A_1\overline{A}_1}+\left(A_1^2\hat{\mathbf{q}}_{2}^{A_1A_1}e^{\text{i}\left(2\omega_{1n} t-2\theta\right)}+c.c.\right),
\end{equation}
with $\hat{\mathbf{q}}_2^{A_1\overline{A}_1}$ and $\hat{\mathbf{q}}_{2}^{A_1A_1}$ computed numerically and displayed in figure~\ref{fig:Fig11}(c)-(d) and (e)-(f), respectively, in terms of second order free surface deformations and potential velocity fields evaluated for $\omega_{1n}=\omega_{11}$. From a numerical perspective, we note that the second order responses can be straightforwardly computed as long as the pairs $\left(\omega,m\right)=\left(2\omega_{1n},2\right)$ and $\left(0,0\right)$ do not correspond to eigenvalues of~\eqref{eq:GenEigProb1}, i.e. the second order operators $\left(\text{i}2\omega_{1n}\mathcal{B}-\mathcal{A}_{2}\right)$ and $-\mathcal{A}_0$ are non-singular and hence invertible.\\
\indent With regards to figure~\ref{fig:Fig11}, it is interesting to note how the second order mean flow potential velocity field is null everywhere. This can be rigorously proven by first noticing that the mean flow corresponds to a time- and azimuthal-averaged flow, i.e. $\partial/\partial t=\partial/\partial \theta=0$. Moreover, in the inviscid limit, free surface elevation and potential field have a $\pi/2$ phase shift, meaning that the first order eigenmode can be normalized such that the eigen-surface is purely real, whereas the eigen-potential is purely imaginary. Under these conditions, the mean flow forcing term on the r.h.s. of the kinematic equation cancels out, so that the associated Laplace equation appears to be constrained by homogeneous Neumann conditions on all the domain boundaries, thus prescribing a trivial constant potential field and therefore a null velocity field. In other words, the second order mean flow system reduces to forced linear meniscus equation (resulting from~\eqref{eq:GovEq_Dyn}) and its conditions at $r=0$ and $r=1$ (both $\partial\hat{\eta}_2^{A_1\overline{A}_1}/\partial r=0$), which prescribes a static mean interface deformation only. Such a result was expected since the second order mean flow response represents the Eulerian mean flow, which, together with the so-called Stokes drift, contribute to the overall Lagrangian mean flow (see \cite{van2018stokes} for a thorough review). While the Stokes drift is a pure kinematic concept, the Eulerian mean flow, often referred to as streaming flow \citep{bouvard2017mean}, is intrinsically a viscous concept. In other words, viscous boundary layers must be reintroduced if one aims to account for it.\\
\indent  We now move forward to the $\epsilon^3$--order problem, which is once again a linear problem forced by combinations of the first and second order solutions as well as by the slow time derivative of the leading order solution and by the external forcing, which was assumed of order $\epsilon^3$,
\begin{eqnarray}
\label{eq:SC_eps3}
\left(\partial_t\mathcal{B}-\mathcal{A}_m\right)\mathbf{q}_3=\mathcal{F}_3=\\
=-\partial_{T_2} A_1\mathcal{B}\hat{\mathbf{q}}_1^{A_1}e^{\text{i}\left(\omega_{1n} t-\theta\right)} + |A_1|^2A_1 \boldsymbol{\hat{\mathcal{F}}}_3^{A_1\overline{A}_1A_1}e^{\text{i}\left(\omega_{1n} t-\theta\right)}+F\boldsymbol{\hat{\mathcal{F}}}_3^F e^{\text{i}\Lambda T_2}e^{\text{i}\left(\omega_{1n} t-\theta\right)}\nonumber \\
+\text{N.R.T.}+c.c.,\nonumber
\end{eqnarray}
with $\boldsymbol{\hat{\mathcal{F}}}_3^F=\left\{0,r/2\right\}^T$ and where $\text{N.R.T.}$ stands for non-resonating terms, which are not relevant for further analysis. As standard in multiple scale analysis, the indeterminacy introduced by the unknown amplitude $A_1$ is resolved by requiring that secular terms do not appear in the solution to equation~\eqref{eq:SC_eps3}. Secularity results from all resonant forcing terms in $\mathcal{F}_3$, i.e. all terms sharing the same frequency and wavenumber $\left(\omega_{1n},1\right)$ of $\mathbf{q}_1$, and in effect all terms explicitly written in~\eqref{eq:SC_eps3}. It follows that a compatibility condition must be enforced through the Fredholm alternative \citep{friedrichs2012spectral}. Such a compatibility condition imposes the amplitude $B=\epsilon A_1e^{\text{i}\lambda t}$ to obey the following normal form 
\begin{equation}
\label{eq:AmpEqSCfinal}
\frac{dB}{dt}=-\text{i}\lambda B + \text{i}\,\mu_{_{SC}} f + \text{i}\,\nu_{_{SC}} |B|^2B,
\end{equation}
where the physical time $t=T_2/\epsilon^2$ has been reintroduced and where forcing amplitude and detuning parameter are recast in terms of their corresponding physical value, $f=\epsilon^3 F$ and $\lambda=\epsilon^2\Lambda$. Moreover, by considering that $\mathbf{q}=\epsilon A_1\hat{\mathbf{q}}_1^{A_1}+\hdots$, the small implicit parameter $\epsilon$ is eliminated by defining the total physical amplitude $A=\epsilon A_1$. The subscript $_{SC}$ stands for single--crest (SC). The various normal form coefficients, which turn out to be real-valued quantities owing to the absence of dissipation, are computed as scalar products between the adjoint mode, $\hat{\mathbf{q}}_1^{A_1 \dagger}$, associated with $\hat{\mathbf{q}}_1^{A_1}$, and the third order resonant forcing terms as follows
\begin{subequations}
\begin{equation}
\label{eq:SC_eps3_coeff_mu}
\text{i}\,\mu_{_{SC}}=\frac{<\hat{\mathbf{q}}_1^{A_1 \dagger},\mathcal{B}\hat{\mathcal{F}}_3^F>}{<\hat{\mathbf{q}}_1^{A_1 \dagger},\mathcal{B}\hat{\mathbf{q}}_1^{A_1}>}=\frac{\int_{z=0}^{}r\overline{\hat{\eta}}_1^{A_1 \dagger}/2\, r\text{d}r}{\int_{z=0}^{}\left(\hat{\eta}_1^{A_1 \dagger}\hat{\Phi}_1^{A_1}+\hat{\Phi}_1^{A_1 \dagger}\hat{\eta}_1^{A_1}\right)\, r\text{d}r},
\end{equation}
\begin{equation}
\label{eq:SC_eps3_coeff_nu}
\text{i}\,\nu_{_{SC}}=\frac{<\hat{\mathbf{q}}_1^{A_1 \dagger},\mathcal{B}\boldsymbol{\hat{\mathcal{F}}}_3^{A_1\overline{A}_1A_1}>}{<\hat{\mathbf{q}}_1^{A_1 \dagger},\mathcal{B}\hat{\mathbf{q}}_1^{A_1}>}=\frac{\int_{z=0}^{}\left(\hat{\eta}_1^{A_1 \dagger}\hat{\mathcal{F}}_{3_{\text{dyn}}}^{A_1\overline{A}_1A_1}+\hat{\Phi}_1^{A_1 \dagger}\hat{\mathcal{F}}_{3_{\text{kin}}}^{A_1\overline{A}_1A_1}\right)\, r\text{d}r}{\int_{z=0}^{}\left(\hat{\eta}_1^{A_1 \dagger}\hat{\Phi}_1^{A_1}+\hat{\Phi}_1^{A_1 \dagger}\hat{\eta}_1^{A_1}\right)\, r\text{d}r}.
\end{equation}
\end{subequations}
Here $\hat{\mathbf{q}}_1^{A_1 \dagger}=\overline{\hat{\mathbf{q}}}_1^{A_1}$, since the inviscid problem is self--adjoint with respect to the Hermitian scalar product $<\mathbf{a},\mathbf{b}>=\int_{\Sigma}^{}\overline{\mathbf{a}}\cdot\mathbf{b}\,\text{d}V$, with $\mathbf{a}$ and $\mathbf{b}$ two generic vector (see \cite{Viola2018a} for a thorough discussion and derivation of the adjoint problem). For the sake of brevity, the explicit expression of $\boldsymbol{\hat{\mathcal{F}}}_3^{A_1\overline{A}_1A_1}$ is omitted, as it only involves straightforward calculations, i.e. a Taylor expansion of nonlinear governing equations and boundary conditions~\eqref{eq:GovEq_Lap}-\eqref{eq:GovEq_Kin} around the rest state $\mathbf{q}_0=\mathbf{0}$. Here we simply denote with the subscript $_{dyn}$ and $_{kin}$ the forcing components appearing in the dynamic and kinematic boundary condition, respectively.\\
\indent By turning to polar coordinates, $B=|B|e^{\text{i}\Theta}$, splitting the modulus and phase parts of~\eqref{eq:AmpEqSCfinal} and looking for stationary solution, $d/dt=0$ with $|B|\ne0$, the following implicit relation is obtained,
\begin{equation}
\label{eq:AmpEqSCfinal_sol}
\tilde{d}_s\Omega^2\mp \frac{\left(\lambda-\nu_{_{SC}}|B|^2\right)|B|}{\mu_{_{SC}}}=0,
\end{equation}
or, in a more common polynomial form,
\begin{equation}
\label{eq:AmpEqSCfinal_sol_poly}
P\left(|B|\right)=|B|^3-\frac{\lambda}{\nu_{_{SC}}}|B|\pm\frac{\mu_{_{SC}}\tilde{d}_s\Omega^2}{\nu_{_{SC}}}=0,
\end{equation}
where $f=\tilde{d}_s\Omega^2$, $\lambda=\Omega-\omega_{1n}$ and the $\mp$ signs correspond to the phases $\Theta=0$ and $\pi$, respectively. The two branches prescribed by~\eqref{eq:AmpEqSCfinal_sol} for $|B|$ as a function of $\Omega$ at a fixed non-dimensional shaking diameter $\tilde{d}_s$ can be easily computed using the Matlab function \textit{fimplicit}. After evaluating the stable and unstable stationary solutions for $|B|$ and $\Theta$, the total single--crest wave solution is reconstructed  as\\ 
\begin{equation}
\label{eq:SC_sol_reconst}
\mathbf{q}_{SC}=\left\{\Phi,\eta\right\}^T=\mathbf{q}_1+\mathbf{q}_2.
\end{equation}
\indent

\subsubsection{Experiments vs weakly nonlinear prediction: wave amplitude}\label{subsubsec:Sec3sub1subsub2}

\begin{figure}
\centering
\includegraphics[width=\textwidth]{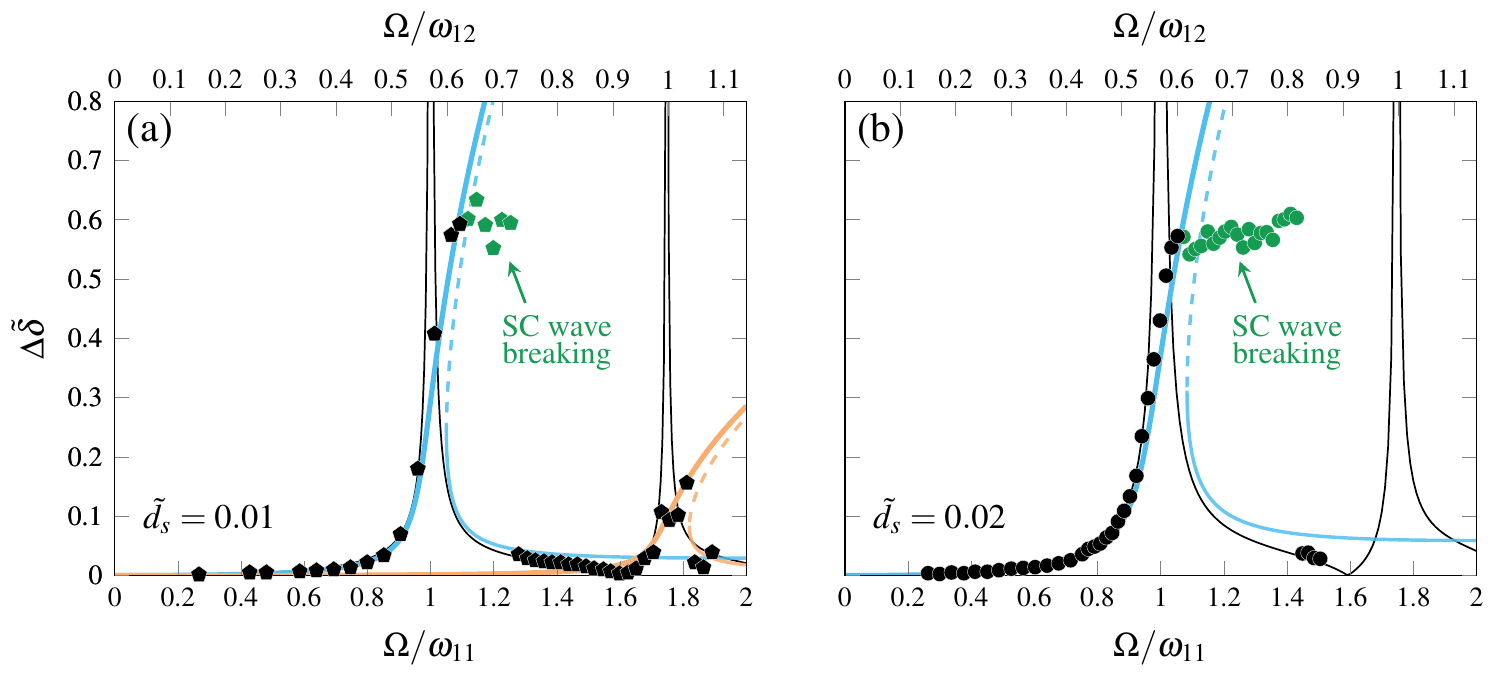}
\caption{Comparison of the WNL-MTS prediction with experiments in terms of maximum crest-to-trough contact line amplitude (non-dimensional), $\Delta\tilde{\delta}$. (a) Black filled pentagons correspond to the experimental measurements presented by \cite{reclari2013hydrodynamics} for $\tilde{d}_s=d_s/D=0.01$, $\tilde{H}=h/D=0.5$ and $D=0.287\,\text{m}$, where the first two non-axisymmetric mode with $\left(m,n\right)=\left(1,1\right)$ and $\left(1,2\right)$ are detected. Solid black lines: solution of the linear potential model according to~\eqref{eq:WNL_Forc_eps1}. Colored lines: WNL-MTS prediction~\eqref{eq:SC_sol_reconst}. The unstable branch is represented as a dashed line. The values of normal form coefficient $\mu_{SC}$ and $\nu_{SC}$ computed for $\Omega\approx\omega_{11}$ and $\omega_{12}$ are given in table~\ref{tab:tab_amp_eq_coeff_SC}. (b) Same as (a) with the black filled circles corresponding to the case of figure~\ref{fig:Fig1} with $\tilde{d}_s=0.02$, $\tilde{H}=0.52$ and $D=0.287\,\text{m}$ and for mode $\left(1,1\right)$, $\Omega\approx\omega_{11}$.}
\label{fig:Fig2} 
\end{figure}

In figure~\ref{fig:Fig2}(a) and (b) the weakly nonlinear (WNL) prediction in terms of maximum crest-to-trough contact line amplitude, $\Delta\tilde{\delta}$, for SC waves is compared with two sets of experimental measurements and with the potential linear solution~\eqref{eq:WNL_Forc_eps1}. In comparison to the linear theory presented in \S\ref{sec:Sec2}, the agreement with experiments improves for different shaking diameters and for different harmonic resonances, e.g. those associated with modes $\left(m,n\right)=\left(1,1\right)$ and $\left(1,2\right)$ of figure~\ref{fig:Fig2}(a). The \textit{hardening} nonlinearity is correctly captured and the amplitude prediction matches well the measurements until the free surface eventually breaks and the wave regime leaves the weakly nonlinear regime, hence suggesting the little relevance of dissipative effects attributable to viscosity in this regime.\\
\indent However, one must note that in this weakly nonlinear approach the driving frequency is essentially fixed around that of a unique non-axisymmetric natural mode, $\Omega\approx\omega_{1n}$. Consequently, when performing the analysis for a mode $\left(1,n\right)$, the influence of all other modes is completely overlooked. 
\begin{table}
\centering
\begin{tabular}{c|c|c|c|cr}
$\left(m,n\right)$ & $\tilde{H}=h/D$ & $D$ $\left[\text{m}\right] $ & $\omega_{mn}$ & $\mu_{SC}$ & $\nu_{SC}$\\ \hline
$\left(1,1\right)$ & 0.50 & 0.287 & $\omega_{11}=\,$1.3239 & 0.27665 & 1.5265\\
$\left(1,2\right)$ & 0.50 & 0.287 & $\omega_{12}=\,$2.3206 & 0.04189 & 17.0246\\
$\left(1,1\right)$ & 0.52 & 0.287 & $\omega_{11}=\,$1.3286 & 0.27761 & 1.4847\\
\end{tabular}
\caption{Value of the amplitude equation coefficient $\mu_{SC}$ and $\nu_{SC}$ in the conditions of figure~\ref{fig:Fig2}.}
\label{tab:tab_amp_eq_coeff_SC}
\end{table}
In consequence, the accuracy of the asymptotic solution rapidly deteriorates moving away from harmonic resonances, when compared to the linear solution~\eqref{eq:WNL_Forc_eps1}, which turns out to be more accurate. This is visible looking at the bottom stable branch in the multi-solution range of figure~\ref{fig:Fig2}(b) or by looking at the driving frequency window $\Omega\in\left[0.7\omega_{12},0.9\omega_{12}\right]$ in figure~\ref{fig:Fig2}(a). In other words, the detuning parameter should be small in order for the present weakly nonlinear analysis, based on a single mode expansion, to hold. On this regard, as no other natural frequencies are encountered for $\Omega<\omega_{11}$, an exception is made for the left branch associated with the harmonic resonance of the first (or fundamental) non-axisymmetric mode, where an excellent agreement of the single mode prediction, comparable to that of the linear solution, lasts until $\Omega\approx0$, i.e. there is no need to employ a leading order multimodal expansion.

\subsubsection{The \textit{Duffing} oscillator analogy}\label{subsubsec:Sec3sub1subsub3}

Mass--spring models are widely employed in several engineering fields, e.g. in aerospace engineering, for the description of close-to-resonance sloshing motions \citep{moiseev1958theory,bauer1966nonlinear,dodge2000new}, where nonlinearities are of crucial importance. The most popular driven nonlinear mass-spring model is that developed by \cite{duffing1918erzwungene}, who added a cubic nonlinear spring deformation (cubic term) to the classical driven harmonic oscillator
\begin{equation}
\label{eq:DuffingEq}
\ddot{x}+2\sigma\dot{x}+ x+c_3 x^3 = p\cos{\Omega t},
\end{equation}
where $\sigma$ is the damping coefficient and where, depending on the sign of $c_3$ the resonance curve bends and the nonlinear resonance frequency shifts, i.e. it decreases for \textit{softening} spring ($c_3 < 0$), whereas it increases for a \textit{hardening} spring ($c_3 > 0$), thus explaining the original observation of Duffing on vibration mechanism. \cite{ockendon1973resonant} showed via asymptotic expansion of the potential flow solution in the neighborhood of a harmonic resonance that for small external forcing amplitudes, sloshing in a two-dimensional rectangular container responds exactly as an undamped Duffing oscillator (with $\sigma=0$). In Appendix~\ref{sec:AppB}, we briefly show that, as expected, the same holds for close-to-harmonic-resonance sloshing in orbital shaken cylindrical containers, whose formal amplitude equation, starting from the full inviscid hydrodynamic system, was derived in \S\ref{subsec:Sec3sub1} (see equation~\eqref{eq:AmpEqSCfinal}). Typically, when the Duffing equation is employed to model close-to-resonance responses in sloshing dynamics and experimental measurements are available, the nonlinear coefficient is often computed by fitting the experimental measurements. Recently, with regards to quasi-two-dimensional rectangular containers laterally excited, \cite{bauerlein2021phase} have carried out careful quantitative comparisons between experiments and theoretical predictions from the damped Duffing equation, showing that their actual sloshing system is remarkably well described by the forced-damped Duffing oscillator. Nevertheless, for increasing wave amplitude responses, experiments deviate from the Duffing solution, which is not capable to predict correctly the phase lag between driving and response, shown to be the key factor for an accurate estimation of the sloshing amplitude of the maximal nonlinear resonance \citep{bauerlein2021phase,cenedese2020conservative}. We note that, by analogy with the undamped Duffing equation, the weakly nonlinear analysis formalized in \S\ref{subsec:Sec3sub1} exacerbates this aspect, since, owing to the lack of dissipation, it can only predict the classic phase lag bounds, $0$ and $\pi$ (see Appendix~\ref{sec:AppA} for further comments on this regard). Nevertheless, one should notice that this intrinsic limitation turns to be unimportant in cases as those of figure~\ref{fig:Fig2}, where for increasing amplitude a wave breaking eventually occurs and the weakly nonlinear theory as well as the Duffing mechanical analogy no longer apply.\\
\indent

\subsection{Double--crest dynamics (DC)}\label{subsec:Sec3sub2}

We now tackle the double--crest (DC) wave response. Its formalization is slightly more subtle, as it requires a new reordering of the small control parameter magnitudes as well as an unusual form of the leading order problem, involving both a homogenous and a particular solution. We remind that the double--crest dynamics in figure~\ref{fig:Fig1} occurs at a driving frequency $\Omega\approx\omega_{21}/2$. Results at the end of this section will be presented for mode $\left(2,1\right)$, for which experiments are available, however, for the sake of generality, we formalize the analysis for any mode $\left(2,n\right)$, i.e. $\Omega=\omega_{2n}/2+\lambda$, where $\lambda$ is the detuning parameter.

\subsubsection{Formalism}\label{subsubsec:Sec3sub2subsub1}

%By analogy with the previous SC problem, the small detuning parameter $\lambda$ and the slow time scale $T_2$ are assumed of order $\epsilon^2$. 
To determine a suitable scaling for the forcing amplitude $f$ and detuning parameter $\lambda$ it is instructive to look at the experimental measurements shown in figure~\ref{fig:Fig1} for $\Omega$ close to $\omega_{21}/2$. One can see that approaching $\Omega\approx\omega_{21}/2$ from lower frequencies, the double--crest wave emerges on the top of a single--crest dynamics, i.e a single--to--double wave transition takes place, with the latter being correctly described by the linear solution, which still behaves well as $\omega_{21}/2$ is far enough from the harmonic resonance occurring at $\omega_{11}$. It follows that the forcing amplitude $f$ and detuning $\lambda$ could be retained at order $\epsilon$ and the first order problem takes the form~\eqref{eq:LinMatForm}, with $\boldsymbol{\mathcal{F}}_1=F\left\{0,r/2\right\}^Te^{\text{i}\left(\Omega t-\theta\right)}+c.c.=F\boldsymbol{\hat{\mathcal{F}}}_1^Fe^{\text{i}\left(\omega_{2n}/2 t-\theta\right)}e^{\text{i}\epsilon\Lambda t}+c.c.\,$, with $f=\epsilon F$ and $\lambda=\epsilon\Lambda$.\\
\indent Furthermore, in \S\ref{sec:Sec2} we have shown how, close enough to the super-harmonic resonance, the divergent behaviour is produced by a second order resonating term, which breaks the straightforward expansion, as $\epsilon^2$--order terms should not become larger than the $\epsilon$--order ones. In the following, this asymptotic breakdown is overcome by assuming that the leading order solution is given by the sum of (i) a particular solution, given by the linear response to the external forcing computed by solving~\eqref{eq:WNL_Forc_sys_eps1}, and (ii) a homogeneous solution, represented by the natural mode $\left(m,n\right)=\left(2,n\right)$, obtained by solving the generalized eigenvalue problem~\eqref{eq:GenEigProb1}, up to an amplitude to be determined at higher orders. The second order resonating term will then require, in the spirit of multiple timescale analysis, an additional second order solvability condition, complementing the third order non-resonance condition already obtained in the single--crest wave weakly nonlinear model. This suggests that two slow time scales exist, namely $T_1$ and $T_2$, with $T_1$ one $\epsilon$--order faster than $T_2$, hence implying that quadratic nonlinearities are stronger than cubic ones. To summarize, the fundamental scalings \textcolor{black}{underpinning} the weakly nonlinear multiple scale expansion for double--crest waves are the following:
\begin{equation}
\label{eq:scaling_DC}
f=\epsilon F,\ \ \ \ \ \lambda=\epsilon\Lambda,\ \ \ \ \ T_1=\epsilon t,\ \ \ \ \ T_2=\epsilon^2 t,
\end{equation}
\begin{equation}
\label{eq:DC_sol_eps1}
\mathbf{q}_{1}=A_2\left(T_1,T_2\right)\hat{\mathbf{q}}_1^{A_2}e^{\text{i}\left(\omega_{2n} t-2\theta\right)}+F\hat{\mathbf{q}}_1^Fe^{\text{i}\left(\left(\omega_{2n}/2\right) t-\theta\right)}e^{\text{i}\Lambda T_1}+c.c.,
\end{equation}
where the unknown slow time amplitude modulation, $A_2$, is here a function of the two time scales $T_1$ and $T_2$, while the amplitude of the particular solution simply equals the forcing amplitude and $\hat{\mathbf{q}}_1^F$ is computed from~\eqref{eq:WNL_Forc_sys_eps1} for $\Omega=\omega_{2n}/2$.\\
\indent The second order linearized forced problem reads
\begin{equation}
\label{eq:DC_eps2}
\left(\partial_t\mathcal{B}-\mathcal{A}_m\right)\mathbf{q}_2=\mathcal{F}_2=\mathcal{F}_2^{i,j}-\frac{\partial A_2}{\partial T_1}\mathcal{B}\hat{\mathbf{q}}_1^{A_2}e^{\text{i}\left(\omega_{2n} t-2\theta\right)} - \text{i}\Lambda F\mathcal{B}\hat{\mathbf{q}}_1^{\Lambda F}e^{\text{i}\left(\left(\omega_{2n}/2\right)t-\theta\right)}e^{\text{i}\Lambda T_1}+c.c.\,.
\end{equation}
The first order solution is made of four different contributions of amplitude $A_2$, $\overline{A}_2$, $F$ and $\overline{F}$, therefore it generates 10 different second order forcing terms, here denoted by $\boldsymbol{\mathcal{F}}_2^{i,j}$, which exhibit a certain frequency and azimuthal periodicity, $\left(\breve{\omega},\breve{m}\right)$.
\begin{table}
\centering
\begin{tabular}{c|cc|cccccc}
 & $\epsilon A_2$ & $\epsilon F$ & $\epsilon^2 A_2A_2$ & $\epsilon^2 \Lambda F$ & $\epsilon^2 A_2\overline{A}_2$ & $\epsilon^2 F\overline{F}$ & $\epsilon^2 A_2F$ & $\epsilon^2 A_2\overline{F}$\\ \hline
$\breve{m}$ & 2 & 1 & 4 & 1 & 0 & 0 & 3 & 1\\
$\breve{\omega}$ & $\omega_{2n}$ & $\omega_{2n}$/2 & $2\omega_{2n}$ & $\omega_{2n}/2$ & 0 & 0 & 3$\omega_{2n}$/2 & $\omega_{2n}$/2\\
\end{tabular}
\caption{First order linear solutions and second order non-resonating forcing terms gathered by their amplitude dependency and corresponding azimuthal and temporal periodicity $\left(\breve{m},\breve{\omega}\right)$. Six terms have been omitted as they are the complex conjugates of $\epsilon A_2$, $\epsilon F$, $\epsilon^2 A_2A_2$, $\epsilon^2 F F$, $\epsilon^2A_2F$ and $\epsilon^2 A_2\overline{F}$.}
\label{tab:Tab1}
\end{table}
The additional two forcing terms stem from the time-derivative of the first order solution~\eqref{eq:DC_sol_eps1} with respect to the first order slow time scale $T_1$. In order to interpret the last term in~\eqref{eq:DC_eps2}, it is worth first noting that, while the amplitude of the linear solution~\eqref{eq:WNL_Forc_eps1}, computed at a generic driving frequency, grows with $\Omega$ as $F/\left(\omega_{11}^2-\Omega^2\right)=\tilde{d}_s\Omega^2/\left(\omega_{11}^2-\Omega^2\right)\propto\Omega^2/\left(\omega_{11}^2-\Omega^2\right)$, in the weakly nonlinear model for double--crest waves, the amplitude of the particular solution~\eqref{eq:DC_sol_eps1} is proportional to $F/\left(\omega_{11}^2-\omega_{21}^2/4\right)=\tilde{d}_s\Omega^2/\left(\omega_{11}^2-\omega_{21}^2/4\right)\sim\Omega^2$, since the driving frequency was frozen at $\Omega=\omega_{21}+\lambda$, with the small detuning parameter, $\lambda$, contributing to modify its phase, but not its amplitude. This leads to an increasing discrepancy between~\eqref{eq:WNL_Forc_eps1} and the leading order particular solution~\eqref{eq:DC_sol_eps1} away from the super-harmonic resonance. The response to the forcing term proportional to $\Lambda F$ in~\eqref{eq:DC_eps2} can be then interpreted as a second order correction of the amplitude of the first order particular solution accounting for a detuning from the exact resonance through $\Lambda F\propto\tilde{d}_s\Omega^2\left(\Omega-\omega_{2n}/2\right)$ and contributing  to improve the asymptotic approximation in a wider range of driving frequency in the neighbourhood of the super-harmonic frequency.\\
\indent None of the forcing terms in~\eqref{eq:DC_eps2} is resonant, as their oscillation frequency and azimuthal wavenumber differ from those of the leading order homogeneous solution, except the term produced by the second--harmonic of the leading order particular solution, i.e. $\boldsymbol{\mathcal{F}}_2^{FF}=F^2\boldsymbol{\hat{\mathcal{F}}}_2^{FF}e^{\text{i}\left(\omega_{2n}-2\theta\right)}e^{\text{i}2\Lambda T_1}+c.c.\,$. To avoid secular terms, a second order compatibility condition is imposed, requiring that the following normal form is verified
\begin{equation}
\label{eq:DC_AmpEq_eps2}
\frac{\partial A_2}{\partial T_1}=\text{i}\,\mu_{_{DC}} F^2e^{\text{i}2\Lambda T_1},
\end{equation}
with $\mu_{_{DC}}$ computed as before, i.e.
\begin{equation}
\label{eq:DC_eps2_coeff_mu}
\text{i}\,\mu_{_{DC}}=\frac{\int_{z=0}^{}\left(\hat{\eta}_1^{A_2 \dagger}\hat{\mathcal{F}}_{2_{\text{dyn}}}^{FF}+\hat{\Phi}_1^{A_2 \dagger}\hat{\mathcal{F}}_{2_{\text{kin}}}^{FF}\right)\, r\text{d}r}{\int_{z=0}^{}\left(\hat{\eta}_1^{A_2 \dagger}\hat{\Phi}_1^{A_2}+\hat{\Phi}_1^{A_2 \dagger}\hat{\eta}_1^{A_2}\right)\, r\text{d}r},
\end{equation}
Taken alone, the dynamics resulting from~\eqref{eq:DC_AmpEq_eps2} is however of little relevance, since the solution, i.e. the frequency-response curve, would still diverge (symmetrically) to infinity for $\Lambda=\Omega-\omega_{2n}/2\rightarrow 0$ in absence of any restoring term, i.e. the nonlinear mechanism responsible for the finite amplitude saturation, which only comes into play at order $\epsilon^3$. This means that the expansion must be pursued up to the following order, and thereby that we must solve for the second-order solution.\\
\indent By substituting~\eqref{eq:DC_AmpEq_eps2} in the forcing expression, equation~\eqref{eq:DC_eps2} can be rearranged as follows
\begin{eqnarray}
\label{eq:DC_eps2_1}
\left(\partial_t\mathcal{B}-\mathcal{A}_m\right)\mathbf{q}_2= \boldsymbol{\mathcal{F}}_{2_{NRT}}^{i,j}+\boldsymbol{\mathcal{F}}_{2_{RT}}^{i,j}=\\
=\boldsymbol{\mathcal{F}}_{2_{NRT}}^{i,j}+F^2\left(\boldsymbol{\hat{\mathcal{F}}}_2^{FF}-\text{i}\,\mu_{_{DC}} \mathcal{B}\hat{\mathbf{q}}_1^{A_2}\right)e^{\text{i}\left(\omega_{2n} t-2\theta\right)}e^{\text{i}2\Lambda T_1}+c.c.\,,\nonumber
\end{eqnarray}
where the subscripts $_{NRT}$ and $_{RT}$ denote non-resonating (whose frequencies and azimuthal periodicities are gathered in table~\ref{tab:Tab1}) and resonating terms, respectively. Note that the term proportional to $\Lambda F$ has been included in the non-resonating forcing terms, whereas the resonant term is written explicitly.
The compatibility condition is now trivially satisfied, meaning that the new forcing term is orthogonal to the adjoint mode, $\hat{\mathbf{q}}_1^{A_2\dagger}=\overline{\hat{\mathbf{q}}}_1^{A_2}$, by construction and therefore, according to the Fredholm alternative, a non-trivial solution exists. Hence, we seek for a second order solution having the following form
\begin{eqnarray}
\label{eq:DC_eps2_full_sol}
\mathbf{q}_2=A_2\overline{A}_2\hat{\mathbf{q}}_2^{A_2\overline{A}_2}+F^2\hat{\mathbf{q}}_2^{F\overline{F}}+\nonumber\\
+A_2^2\hat{\mathbf{q}}_2^{A_2A_2}e^{\text{i}\left(2\omega_{2n}t-4\theta\right)}+\Lambda F\hat{\mathbf{q}}_2^{\Lambda F}e^{\text{i}\left(\left(\omega_{2n}/2\right)t-\theta\right)}e^{\text{i}\Lambda T_1}+\nonumber\\
+A_2F\hat{\mathbf{q}}_2^{A_2F}e^{\text{i}\left(\left(3\omega_{2n}/2\right)t-3\theta\right)}e^{\text{i}\Lambda T_1}+A_2\overline{F}\hat{\mathbf{q}}_2^{A_2\overline{F}}e^{\text{i}\left(\left(\omega_{2n}/2\right)t-\theta\right)}e^{-\text{i}\Lambda T_1}+\nonumber\\
+F^2\hat{\mathbf{q}}_2^{FF}e^{\text{i}\left(\omega_{2n}t-2\theta\right)}e^{\text{i}2\Lambda T_1}+c.c.\,.
\end{eqnarray}
\begin{figure}
\centering
\includegraphics[width=0.9\textwidth]{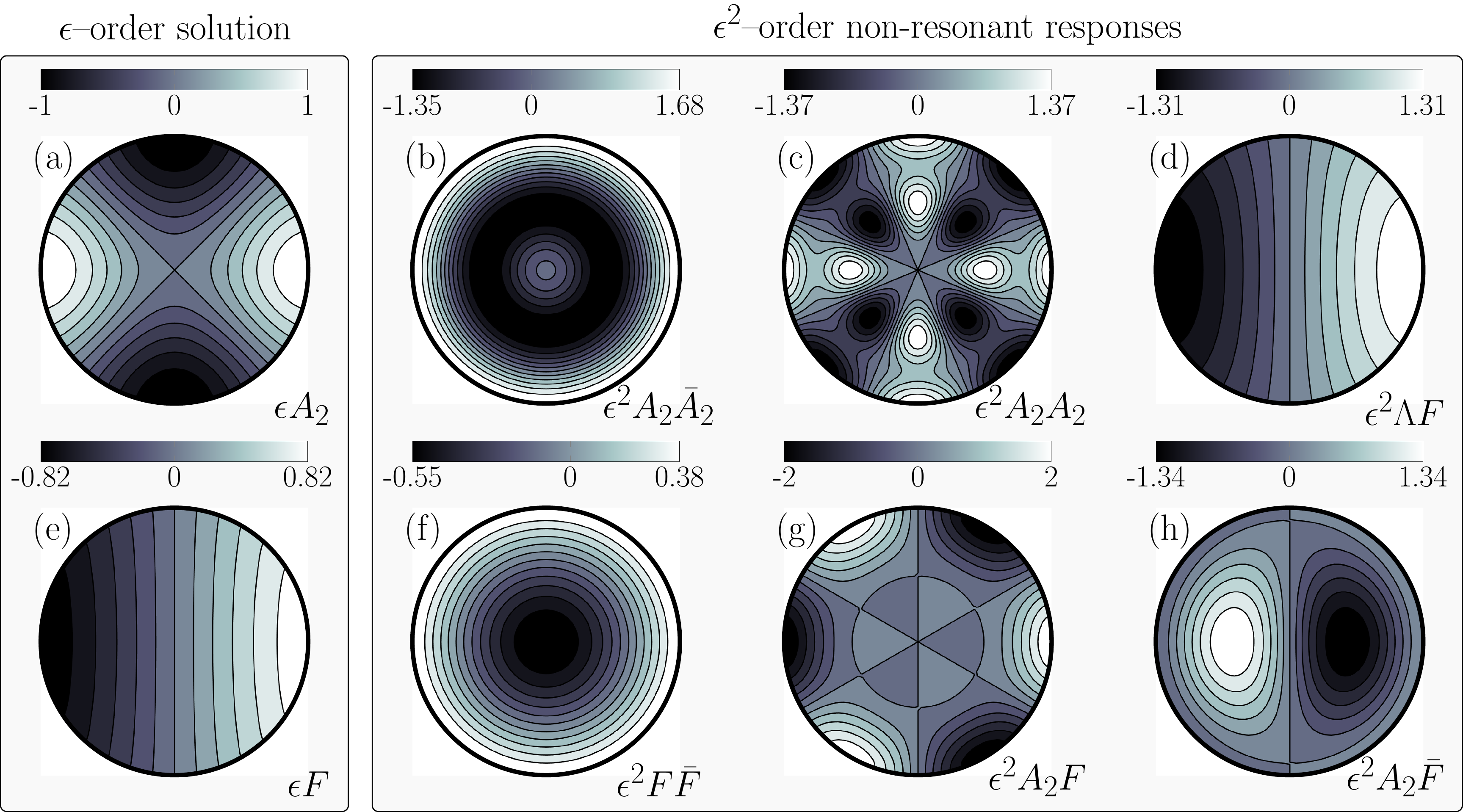}
\caption{Real part of the first, (a) $\hat{\eta}_1^{A_2}$ and (e) $\hat{\eta}_1^{F}$, and non-resonating second order, (b) $\hat{\eta}_2^{A_1\overline{A}_2}$, (c) $\hat{\eta}_2^{A_2A_2}$, (d) $\hat{\xi}_2^{\Lambda F}$, (f) $\hat{\eta}_2^{F\overline{F}}$, (g) $\hat{\eta}_2^{A_2F}$ and (h) $\hat{\eta}_2^{A_2\overline{F}}$, free surface deformations computed for $\omega_{2n}=\omega_{21}$. The first order eigenmode is normalized with the amplitude and phase of the contact line (at $r=1$), such that the free surface, $\hat{\eta}_1^{A_2}$ is purely real, whereas $\hat{\Phi}_1^{A_2}$ is purely imaginary. Note that the second order mean flow $\hat{\eta}_2^{F\overline{F}}$ constantly induces an upside down \textit{bell}-like an axisymmetric interface deformation pushing the free surface downward at the center of the moving container, by analogy with the effect produced by $\hat{\eta}_2^{A_1\overline{A}_1}$ for SC waves, as the two responses are essentially equivalent up to a pre factor. Here the mean flow $\hat{\eta}_2^{A_2\overline{A_2}}$ for DC pushes the interface upward at the wall (same as $\hat{\eta}_2^{F\overline{F}}$) and, at the same time, downward in an annular region at intermediate radial coordinates, without altering the free surface elevation at the container revolutions axis.}
\label{fig:Fig10} 
\end{figure} 
All non-resonant responses in~\eqref{eq:DC_eps2_full_sol} are handled similarly, i.e. they are computed in Matlab by performing a simple matrix inversion using standard LU solvers (as in \S\S\ref{sec:Sec2} and \ref{subsec:Sec3sub1}). As anticipated above, although the operator associated with the resonant forcing term, i.e. $\left(\text{i}\omega_{2n}\mathcal{B}-\mathcal{A}_{2}\right)$, is singular, the value of the normal form coefficient~\eqref{eq:DC_eps2_coeff_mu} ensures that a non-trivial solution for $\hat{\mathbf{q}}_2^{FF}$ exists. Diverse approaches can be followed to compute this response. Here such a response is computed by using the \textit{pseudoinverse} matrix of the singular operator \citep{orchini2016weakly}. Another possible approach is given in Appendix~A of \cite{Meliga2012}, where a two-step regularization procedure, involving an intermediate factious solution for $\hat{\mathbf{q}}_2^{FF}$ is employed. We also note that in~\eqref{eq:DC_eps2_full_sol}, exactly as in~\eqref{eq:SC_eps2}, a second order homogeneous solution has not been accounted for as its introduction would be irrelevant to the final solution.\\
\indent The first order solutions together with all the non-resonating second order responses are shown in the various panels of figure~\ref{fig:Fig10}, where the two leading order contributions, $\epsilon A_2$ and $\epsilon F$, corresponding to the double-- and single--crest wave, respectively, can be identified. Moreover, we note that the second order response proportional to $\epsilon^2\Lambda F$ has a spatial structure similar to that of the leading order response $\epsilon F$, as it represents the second order correction to the latter caused by small frequency shifts of order $\epsilon$.\\
\indent Lastly, at third order in $\epsilon$, the problem reads
\begin{eqnarray}
\label{eq:DC_eps3}
\left(\partial_t\mathcal{B}-\mathcal{A}_m\right)\mathbf{q}_3=\boldsymbol{\mathcal{F}}_3=\\
=-\frac{\partial A_2}{\partial T_2 }\mathcal{B}\hat{\mathbf{q}}_1^{A_2}e^{\text{i}\left(\omega_{2n} t-2\theta\right)} - \text{i}2\Lambda F^2\mathcal{B}\hat{\mathbf{q}}_2^{FF}e^{\text{i}\left(\omega_{2n}-2\theta\right)}e^{\text{i}2\Lambda T_1}+\nonumber\\
+ |A_2|^2A_2 \boldsymbol{\hat{\mathcal{F}}}_3^{A_2\overline{A}_2A_2}e^{\text{i}\left(\omega_{2n} t-2\theta\right)}+A_2F^2\boldsymbol{\hat{\mathcal{F}}}_3^{A_2F\overline{F}}e^{\text{i}\left(\omega_{2n} t-2\theta\right)}+\nonumber\\
+\Lambda F^2\boldsymbol{\hat{\mathcal{F}}}_3^{\Lambda FF}e^{\text{i}\left(\omega_{2n}-2\theta\right)}e^{\text{i}2\Lambda T_1}+\text{N.R.T.}+c.c.\,,\nonumber
\end{eqnarray}
where the first two forcing terms arise from the time-derivative of the first order solution with respect to the second order slow time scale $T_2$ and from that of the second order solution with respect to the first order slow time scale $T_1$, respectively. By noticing that the second and last forcing terms share the same amplitude dependence, i.e. $\Lambda F^2$, they can be recast into a single forcing term, say $\Lambda F^2\hat{\mathcal{F}}_3^{\Lambda FF}e^{\text{i}\left(\omega_{2n}-2\theta\right)}e^{\text{i}2\Lambda T_1}+c.c.\,$.\\
\indent Once again, all terms explicitly written in~\eqref{eq:DC_eps3} are resonant, as they share the same pair $\left(\omega_{2n},2\right)$ than the first order homogeneous solution, hence a third order compatibility condition, leading to the following normal form, must be enforced
\begin{equation}
\label{eq:AmpEqDCfinal0}
\frac{\partial A_2}{\partial T_2}=\text{i}\zeta_{_{DC}} \Lambda F^2e^{\text{i}2\Lambda T_1}+\text{i}\,\chi_{_{DC}}A_2F^2+\text{i}\,\nu_{_{DC}} |A_2|^2A_2,
\end{equation}
with 
\begin{subequations}
\begin{equation}
\label{eq:DC_eps3_coeff_zeta}
\text{i}\,\zeta_{_{DC}}=\frac{\int_{z=0}^{}\left(\hat{\eta}_1^{A_2 \dagger}\hat{\mathcal{F}}_{3_{\text{dyn}}}^{\Lambda FF}+\hat{\Phi}_1^{A_2 \dagger}\hat{\mathcal{F}}_{3_{\text{kin}}}^{\Lambda FF}\right)\, r\text{d}r}{\int_{z=0}^{}\left(\hat{\eta}_1^{A_2 \dagger}\hat{\Phi}_1^{A_2}+\hat{\Phi}_1^{A_2 \dagger}\hat{\eta}_1^{A_2}\right)\, r\text{d}r},
\end{equation}
\begin{equation}
\label{eq:DC_eps3_coeff_chi}
\text{i}\,\chi_{_{DC}}=\frac{\int_{z=0}^{}\left(\hat{\eta}_1^{A_2 \dagger}\hat{\mathcal{F}}_{3_{\text{dyn}}}^{A_2F\overline{F}}+\hat{\Phi}_1^{A_2 \dagger}\hat{\mathcal{F}}_{3_{\text{kin}}}^{A_2F\overline{F}}\right)\, r\text{d}r}{\int_{z=0}^{}\left(\hat{\eta}_1^{A_2 \dagger}\hat{\Phi}_1^{A_2}+\hat{\Phi}_1^{A_2 \dagger}\hat{\eta}_1^{A_2}\right)\, r\text{d}r},
\end{equation}
\begin{equation}
\label{eq:DC_eps3_coeff_nu}
\text{i}\,\nu_{_{DC}}=\frac{\int_{z=0}^{}\left(\hat{\eta}_1^{A_2 \dagger}\hat{\mathcal{F}}_{3_{\text{dyn}}}^{A_2\overline{A}_2A_2}+\hat{\Phi}_1^{A_2 \dagger}\hat{\mathcal{F}}_{3_{\text{kin}}}^{A_2\overline{A}_2A_2}\right)\, r\text{d}r}{\int_{z=0}^{}\left(\hat{\eta}_1^{A_2 \dagger}\hat{\Phi}_1^{A_2}+\hat{\Phi}_1^{A_2 \dagger}\hat{\eta}_1^{A_2}\right)\, r\text{d}r}.
\end{equation}
\end{subequations}
As a last step in the derivation of the final amplitude equation for the double--crest (DC) waves and in order to eliminate the implicit small parameter $\epsilon$, we unify~\eqref{eq:DC_AmpEq_eps2} and~\eqref{eq:AmpEqDCfinal0} into a single equation recast in terms of the physical time $t=T_1/\epsilon=T_2/\epsilon^2$, physical forcing control parameters, $f=\epsilon F$ and $\lambda=\epsilon\Lambda$, and total amplitude, $A=\epsilon A_2$. This is achieved by summing~\eqref{eq:DC_AmpEq_eps2} and~\eqref{eq:AmpEqDCfinal0} along with their respective weights $\epsilon^2$ and $\epsilon^3$, thus obtaining
\begin{equation}
\label{eq:AmpEqDCfinal}
\frac{dB}{dt}=-\text{i}\,\left(2\lambda-\chi_{_{DC}} f^2\right) B + \text{i}\,\left(\zeta_{_{DC}}\lambda + \mu_{_{DC}}\right) f^2 + \text{i}\,\nu_{_{DC}} |B|^2B,
\end{equation}
where the change of variable $A=Be^{\text{i}2 \lambda t}$ has been introduced for convenience. As in \S\ref{subsec:Sec3sub1}, by turning to polar coordinates, $B=|B|e^{\text{i}\Theta}$, splitting the modulus and phase parts of~\eqref{eq:AmpEqDCfinal} and looking for stationary solution, $d/dt=0$ with $|B|\ne0$, the following implicit relation is obtained,
\begin{equation}
\label{eq:AmpEqDCfinal_sol}
\tilde{d}_s\Omega^2-\sqrt{\left(2\lambda-\nu_{_{DC}}|B|^2\right)|B|/\left[\chi_{_{DC}}|B|\pm \left(\zeta_{_{DC}}\lambda+\mu_{_{DC}}\right)\right]}=0,
\end{equation}
\indent where only the real solutions corresponding to $f=\tilde{d}_s\Omega^2>0$ are retained, as the combinations $\tilde{d}_s\Omega^2<0$ are not physically meaningful.\\
Although two more terms appear in~\eqref{eq:AmpEqDCfinal} and the dependence on the forcing amplitude is different with respect to the SC case, i.e. $f^2$ instead of $f$, thus leading to the square root in~\eqref{eq:AmpEqDCfinal_sol}, amplitude equation~\eqref{eq:AmpEqDCfinal} is reminiscent of that given in~\eqref{eq:AmpEqSCfinal}. Indeed, equation~\eqref{eq:AmpEqDCfinal} contains essentially three contributions, 
\begin{equation}
\label{eq:analogy_SC_DC}
\lambda \leftrightarrow \left(2\lambda-\chi_{_{DC}}f^2\right),\ \ \ \mu_{_{SC}}f\leftrightarrow\left(\zeta_{_{DC}}\lambda+\mu_{_{DC}}\right)f^2,\ \ \ \nu_{_{SC}}\leftrightarrow\nu_{_{DC}},
\end{equation}
in order, a detuning term (forcing amplitude dependent), an additive (quadratic) forcing term (forcing frequency dependent) and the classic cubic restoring term, respectively. Hence, the same qualitative \textit{hardening} or \textit{softening} nonlinear behaviours as well as hysteresis, typical features of the Duffing-equation, are expected under the hypotheses of the present analysis.\\
\indent The total flow solution predicted by the WNL for DC waves and reconstructed as
\begin{equation}
\label{eq:DC_sol_reconst}
\mathbf{q}_{DC}=\left\{\Phi,\eta\right\}^T=\mathbf{q}_1+\mathbf{q}_2,
\end{equation}
is compared in figures~\ref{fig:Fig3} and~\ref{fig:Fig31} with experiments from \cite{reclari2013hydrodynamics} and \cite{reclari2014surface} (see also figure~\ref{fig:Fig1}).

\begin{figure}
\centering
\includegraphics[width=0.9625\textwidth]{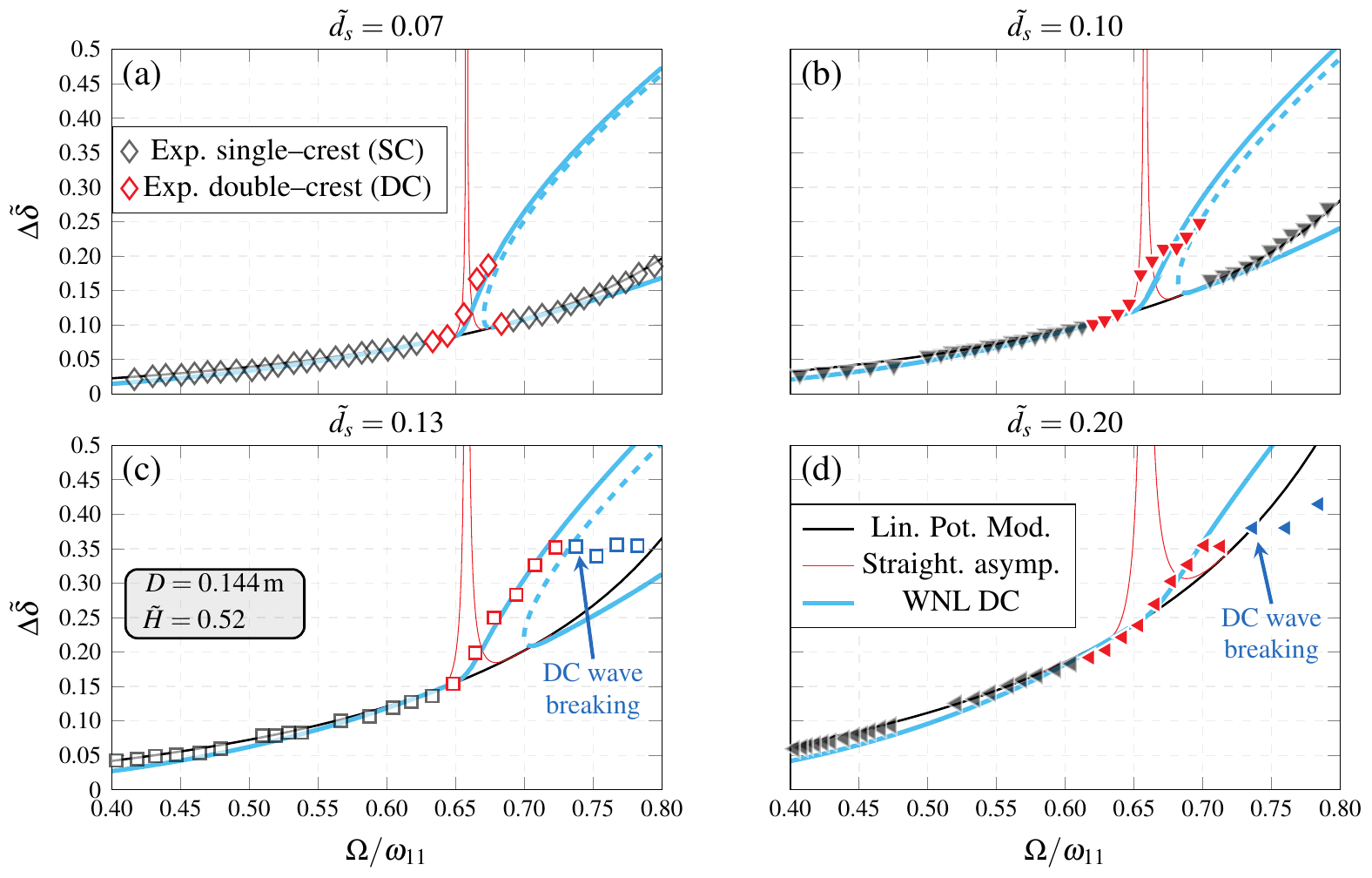}
\caption{Weakly nonlinear (WNL) prediction for double--crest (DC) waves versus experiments by \cite{reclari2014surface} (reported in figure~\ref{fig:Fig1} of this paper) in terms $\Delta\tilde{\delta}$ for the smallest container diameter $D=0.144\,\text{m}$, for different shaking diameters and at a driving frequency close to $\Omega\approx\omega_{21}/2$ ($\Omega/\omega_{11}=0.6576$). Solid black lines: linear potential solution~\eqref{eq:WNL_Forc_eps1}. Red solid lines: straightforward asymptotic solution~\eqref{eq:WNLS_Forc_Tot}. Light blue solid and dashed lines: stable and unstable branches, respectively, predicted by the WNL via~\eqref{eq:AmpEqDCfinal_sol}. The normal form coefficient values for this configurations are $\chi_{_{DC}}=3.4572$, $\zeta_{_{DC}}=0.8708$, $\mu_{_{DC}}=0.1292$ and $\nu_{_{DC}}=10.0181$.}
\label{fig:Fig3} 
\end{figure}
\begin{figure}
\centering
\includegraphics[width=0.9625\textwidth]{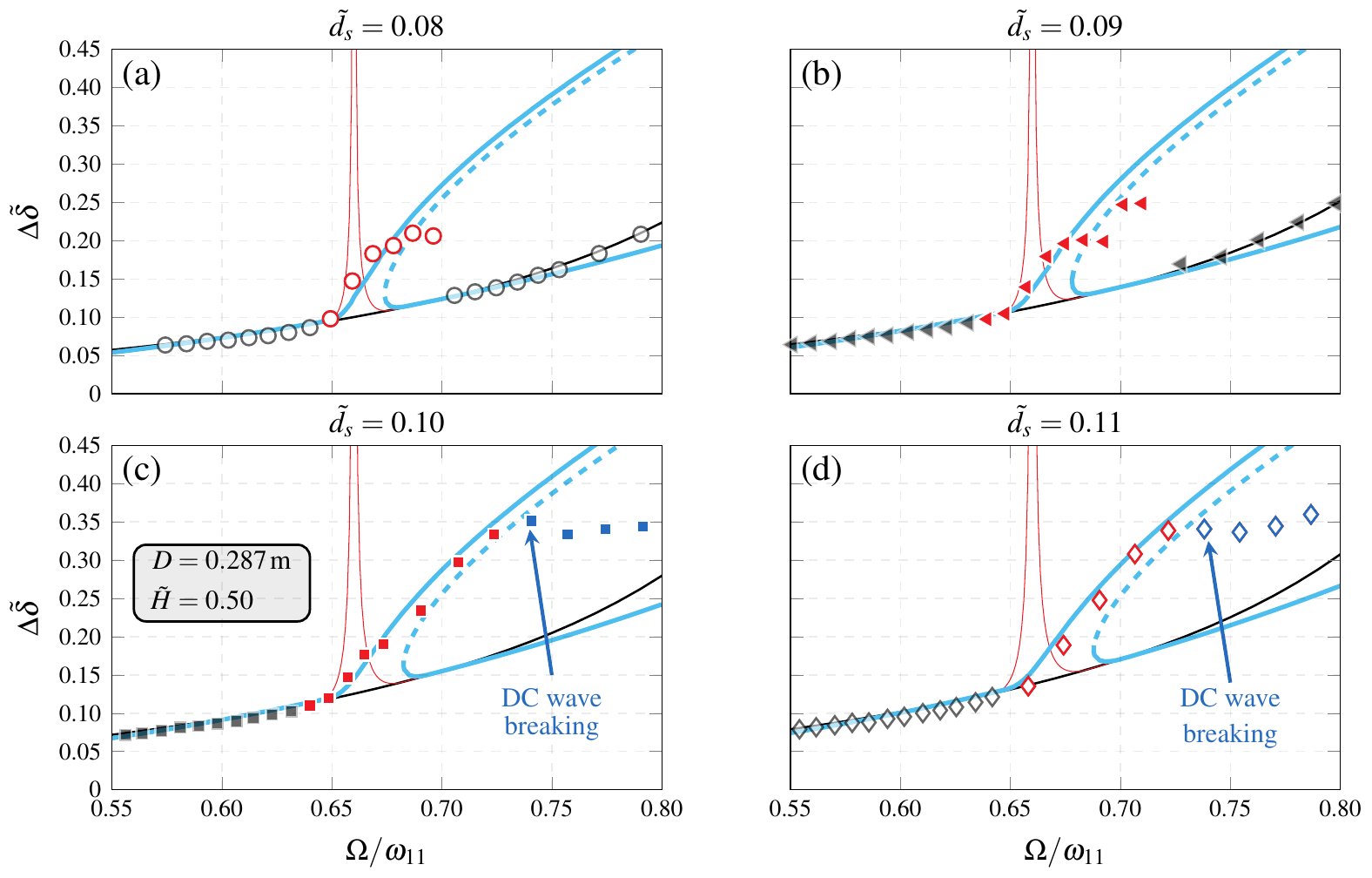}
\caption{Same as figure~\ref{fig:Fig3} but for the largest container diameter $D=0.287\,\text{m}$ and $\tilde{H}=0.50$ (see figure~4.19 of \cite{reclari2013hydrodynamics}). The normal form coefficient values for this configurations are $\chi_{_{DC}}=3.4718$, $\zeta_{_{DC}}=0.9059$, $\mu_{_{DC}}=0.1338$ and $\nu_{_{DC}}=9.8847$.}
\label{fig:Fig31} 
\end{figure}

\subsubsection{Experiments vs weakly nonlinear prediction: wave amplitude}\label{subsubsec:Sec3sub2subsub2}

In figures~\ref{fig:Fig3} and~\ref{fig:Fig31}, the weakly nonlinear (WNL) prediction of double--crest (DC) waves is quantitatively compared with the experimental measurements from \cite{reclari2013hydrodynamics} and \cite{reclari2014surface} in terms of maximum non-dimensional crest-to-trough contact line amplitude, $\Delta\tilde{\delta}$, for different values of the shaking diameters, $\tilde{d}_s$ corresponding to those of figure~\ref{fig:Fig1} in the frequency window close to $\omega_{21}/2$.\\
\indent The improvement gained through the formal WNL analysis, when compared with the linear and straightforward asymptotic models, is striking. The amplitude equation model correctly predicts the transition from a single-- to a double--crest wave and the resulting finite amplitude saturation via \textit{hardening} nonlinear mechanism, thus remarkably narrowing the gap with experiments for all the values of $\tilde{d}_s$ considered and for different container configurations.\\
\indent Notwithstanding such an improvement, figure~\ref{fig:Fig3} highlights the main limitation of the present amplitude equation model for DC waves. Indeed, one notices that, while at larger shaking diameters, i.e. $\tilde{d}_s=0.13$ and $0.20$, a DC wave first emerges on the top a single--crest (SC) dynamics and eventually a double--crest wave breaking occurs at larger frequencies, a jump-down transition from DC to SC takes place by increasing $\Omega$ at lower shaking diameters, i.e. $\tilde{d}_s=0.07$ and $0.10$ for $D=0.144\,\text{m}$. This well-known hysteretic behaviour can be reasonably ascribed to the viscous dissipation of the system. For instance, at sufficiently small shaking diameters, e.g. $\tilde{d}_s\approx0.02$ (see figure~\ref{fig:Fig1}), the DC dynamics does not manifest at all, as the energy pumped into the system by the external forcing is likely not sufficient to dominate over the system viscous dissipation, whose effect also depends on the container diameter, $D$. Indeed, figure~\ref{fig:Fig31} clearly shows that larger diameters, i.e. $D=0.287\,\text{m}$, generate less dissipation. It follows that for larger $D$, by increasing the driving frequency at a fixed shaking diameter, e.g. $\tilde{d}_s=0.10$, the free surface is more likely to undergo a wave breaking, rather than a jump-down transition (see figures~\ref{fig:Fig3}(b) and~\ref{fig:Fig31}(c)). Obviously, the inviscid model employed here is not capable to predict the so-called jump-down frequency. In Appendix~\ref{sec:AppA}, a heuristic viscous damping model is introduced to tentatively overcome the beforehand mentioned limitations.\\
\indent Finally, we note that, for frequency moderately far from the super-harmonic resonance, the agreement of the WNL model with experiments and with the linear solution, which behaves well away from $\Omega\approx\omega_{11}$, progressively deteriorates. This is particularly evident on the right stable branch and can be ascribed to the fact that the asymptotic model is essentially formalized for a fixed driving frequency, i.e. $\Omega\approx\omega_{21}/2$, thus filtering out the existence of other resonances. Nevertheless, owing to the assumption of detuning parameter of order $\epsilon$, the second order correction to the leading order particular solution guarantees a fairly good agreement in a relatively wide range of frequency around $\omega_{21}/2$.

\begin{figure}
\centering
\includegraphics[width=0.85\textwidth]{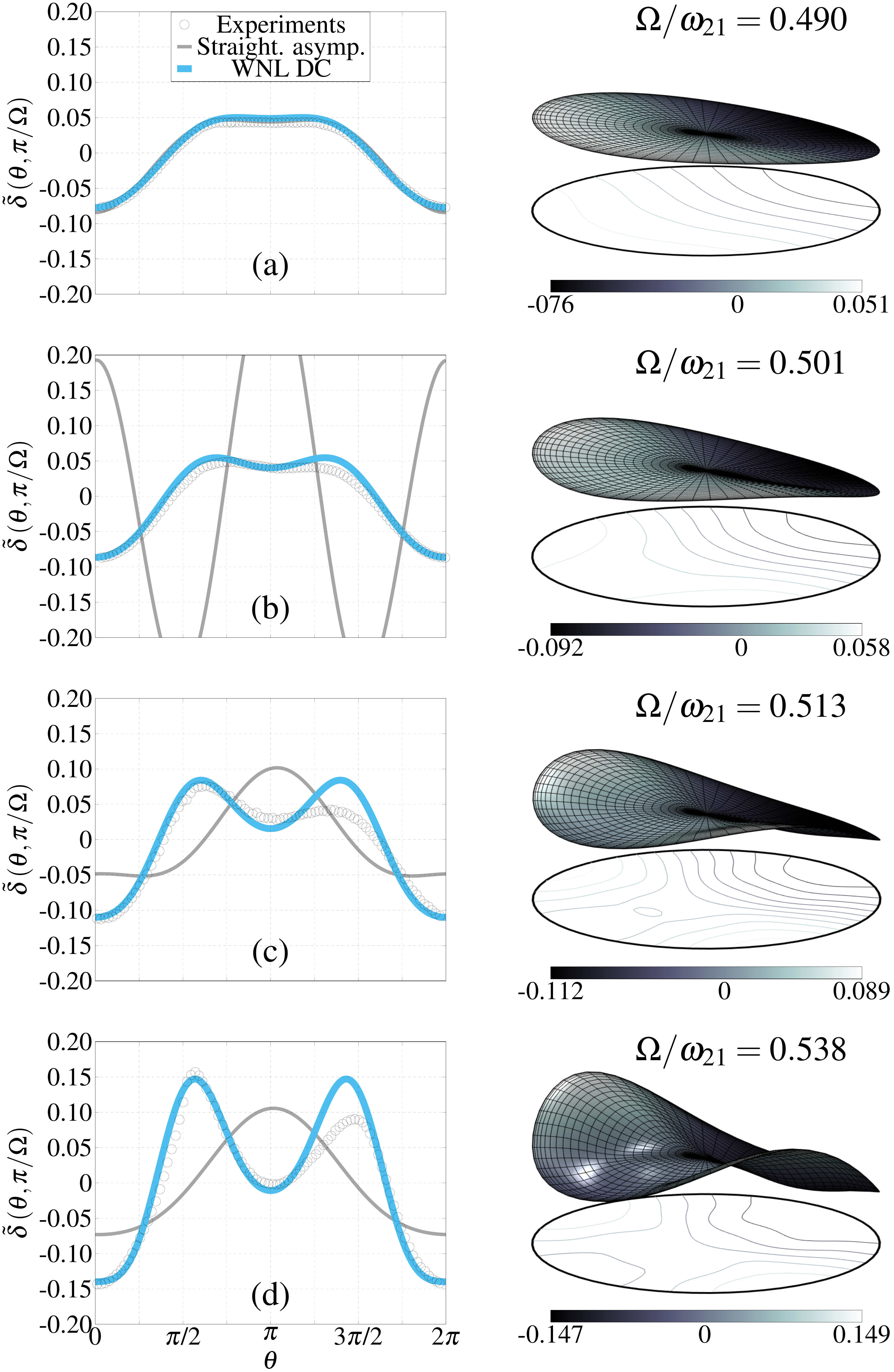}
\caption{Left-panels: comparison of the dimensionless and phase-averaged wave height measured at the wall, $\tilde{\delta}\left(\theta,\pi/\Omega\right)$, (black circles) with the straightforward asymptotic solution rebuilt via (3.14) (gray solid line) and the weakly nonlinear (WNL) solution for the double–crest (DC) wave (4.25). Panels correspond to $\tilde{H}=0.52$, $\tilde{d}_s=0.11$ and $D=0.144\,\text{m}$. The experimental measurements, here shown as black circles, are available in \cite{reclari2013hydrodynamics}, except for panel (c), which is provided in \cite{reclari2014surface}. Note that (b) the nonlinear prediction has a very large amplitude. Right-panels: corresponding three-dimensional free surface deformation, $\eta\left(r,\theta,\pi/\Omega\right)$, reconstruct via (4.25). The single–to–double crest transition via hardening nonlinearity is clearly visible moving from top to bottom, i.e. for increasing frequency.}
\label{fig:Fig4} 
\end{figure}

\subsubsection{Experiments vs weakly nonlinear prediction: free surface reconstruction}\label{subsubsec:Sec3sub2subsub3}

In figure~\ref{fig:Fig4}, the weakly nonlinear (WNL) models for the double--crest waves (DC) is compared versus the straightforward asymptotic prediction discussed in \S\ref{sec:Sec2} and the experimental measurements for DC waves from \cite{reclari2013hydrodynamics} and \cite{reclari2014surface}. The direct quantitative comparison is here outlined in terms of dimensionless and phase-averaged wave height measured at the sidewall, $\tilde{\delta}\left(\theta\right)$. \\
\indent We observe that, if at $\Omega/\omega_{21}=0.490$ both models match satisfactorily the experimental points, as soon as $\Omega/\omega_{21}=0.5$ is approached, the straightforward asymptotic solution diverges due to the resonant (second order) super-harmonic term, while the WNL solution predicts correctly the finite amplitude saturation and the emergence of a DC wave on the top of a single--crest (SC) one. The WNL model for DC waves remains in fairly good agreement even at larger driving frequency, although the increasing phase-asymmetry between the two local peaks at $\theta=\pi/2$ and $3\pi/2$ is not retrieved by the present inviscid asymptotic analysis, where secondary effects, e.g. the phase shift induced by viscous dissipation and influence of other higher modes, as well as stronger nonlinear effects for increasing wave amplitudes are overlooked.\\
\indent For completeness, the three-dimensional free surface, $\eta\left(r,\theta,\pi/\Omega\right)$, is reconstructed through~\eqref{eq:DC_sol_reconst} and shown in the right-panels of figure~\ref{fig:Fig4}, where, for increasing shaking frequencies, the nonlinear transition from a nearly single--crest wave dynamics to a double--crest wave dynamics is enlightened.

%\indent Let us first focus on the single--crest wave dynamics shown in figure.~\ref{fig:Fig4}-(a), (b), (c) and (d). We observe that at driving frequencies far enough and lower than $\Omega/\omega_{11}=1$, the straightforward-WNL solution~\eqref{eq:WNLS_Forc_Tot}, which sufficiently away from $\Omega/\omega_{21}=0.5$ is essentially governed by the first order linear solution~\eqref{eq:WNL_Forc_eps1}, is slightly more accurate than the WNL-MTS for SC waves and it is in fact undistinguishable from the experimental measurements. This is again explainable by noticing that, whereas the accuracy of the linear solution strongly increase far from any resonances (see figure.~\ref{fig:Fig1}), that of the WNL-MTS model, where the driving frequency is frozen around a specific value (except for a small detuning parameter), decrease when shifting away from that $\Omega$ value. Nevertheless, as $\Omega/\omega_{11}\approx1$ is approached, the straightforward--WNL solution tends to diverge, while the WNL-MTS model predicts a correct wave steepening, hence closing the gap with experiments and providing a notable agreement.\\

\subsubsection{The \textit{Helmholtz}--\textit{Duffing} oscillator analogy}\label{subsubsec:Sec3sub2subsub4}

While the Duffing equation is known to capture period--3 and period--1/3 dynamics arising from the cubic nonlinearity \citep{jordan1999nonlinear,kalmar2011forced}, as those observed by \cite{bauerlein2021phase} and occasionally by \cite{reclari2014surface}, it cannot predict the period--halving dynamics associated with the super-harmonic resonance investigated in this paper. Therefore, in connection with \S\ref{subsubsec:Sec3sub1subsub3}, here we aim to identify the simplest possible mechanical oscillator that could mimic, at least from a qualitative perspective, the period-1/2 dynamics studied in this work.\\
\indent The weakly nonlinear analysis (WNL) as well as the straightforward asymptotic model highlighted the crucial role of quadratic nonlinearities emerging at second order and from which the double--crest (DC) dynamics stems. At the same time, the WNL model enlightened that second order terms only are not sufficient to capture all the dynamics features owing to the lack of restoring terms and, therefore, cubic nonlinearities must be retained. These considerations suggest that the DC dynamics could be tentatively described by a driven oscillator with both quadratic (asymmetric) and cubic (symmetric) nonlinear terms, i.e.
\begin{equation}
\label{eq:HD_Eq}
\ddot{x}+2\sigma\dot{x}+ x+c_2 x^2 + c_3 x^3 = p\cos{\Omega t}.
\end{equation}
Equation~\eqref{eq:HD_Eq}, also commonly known as Helmholtz--Duffing equation, has wide applications in engineering problems as those related to beams, plates and shells subjected to an initial static curvature \citep{mirzabeigy2014approximate,askari2011approximate}, whose governing equations are reconduced to a second-order nonlinear ordinary equation with quadratic and cubic nonlinear terms \citep{ke2010analytical,alijani2011nonlinear,fallah2011nonlinear}.\\
\indent Among the diverse asymptotic solutions of~\eqref{eq:HD_Eq} in different limits \citep{rega1995bifurcation,benedettini1989planar,kovacic2011duffing}, the most relevant to our work is that of \cite{benedettini1989planar}. Within the context of planar nonlinear response of suspended elastic cables to an external excitation, they derived an amplitude equation which concerns with the first or fundamental super-harmonic excitation, i.e. $\Omega\approx1/2$, of~\eqref{eq:HD_Eq}. Their weakly nonlinear approach is detailed in Appendix~\ref{sec:AppC}, with the additional assumption of vanishing damping $\sigma=0$. Assuming $2\Omega=1+\lambda=1+\epsilon\hat{\lambda}$, small nonlinearities, $c_2=\epsilon\hat{c_2}$ and $c_3=\epsilon^2\hat{c}_3$, and introducing two slow time scales, one obtains
\begin{equation}
\label{eq:HD_Eq_AmpEq}
dD/dt=-\text{i}\,\left(\lambda +c_5f^2 \right)D +\text{i}\,\left(1-\lambda/2\right)c_2f^2/2-\text{i}\,4c_4|D|^2D,
\end{equation}
with $C=De^{\text{i}\lambda t}$ and with the auxiliary coefficients $c_4$ and $c_5$ (both functions of $c_2$ and $c_3$) defined in \cite{benedettini1989planar}. By comparing term by term, the analogy with equation~\eqref{eq:AmpEqDCfinal} is evident.\\
\indent To conclude, although the DC dynamics examined in this paper is intrinsically related to the simultaneous interplay of multiple waves, thus making particularly challenging an accurate \textit{vis}-\textit{à}-\text{vis} quantitative comparison with a single-degree-of-freedom mechanical model, equation~\eqref{eq:HD_Eq_AmpEq} seems to suggest that the actual inviscid sloshing dynamics in the DC regime may be, at least qualitatively, described by the undamped Helmholtz--Duffing equation~\eqref{eq:HD_Eq} driven super-harmonically.

%\clearpage
\section{Conclusion}\label{sec:Conc}

With regards to orbital shaken cylindrical containers and, specifically, to the careful experimental campaign reported in \cite{reclari2013hydrodynamics} and \cite{reclari2014surface}, a weakly nonlinear analysis (WNL) via multiple timescale method was formalized in \S\ref{sec:Sec3} in order to investigate diverse features of the steady state free surface dynamics and, particularly, the double--crest (DC) wave dynamics pertaining at half the frequency of the first $m=2$ natural mode.\\
\indent After having discussed the substantial limitations of the straightforward expansion procedure propose by \cite{reclari2014surface} and summarized in \S\ref{sec:Sec2}, the WNL analysis was first formulated under the most common condition of pure harmonic resonance. Despite the inviscid assumption, the WNL analysis developed for the single--crest (SC) wave dynamics was shown to be in fairly good agreement with all the experimental measurements. In fact, the present model correctly describes the close-to-resonance \textit{hardening} nonlinear behaviour experimentally observed. The agreement remains sufficiently accurate until the free surface eventually breaks and a transition to a fully nonlinear regime occurs.\\
\indent It is well-assessed in the literature that the close-to-harmonic-resonance sloshing dynamics can be modeled (from both qualitative and quantitative perspectives \citep{bauerlein2021phase}) by a single degree of freedom (1dof) system with a cubic nonlinearity and driven harmonically, i.e. by the famous Duffing oscillator, as rigorously proved for a two-dimensional rectangular container laterally excited \citep{ockendon1973resonant}. Without surprise, this was shown to hold for the case of orbital shaken cylindrical containers as well.\\
\indent The WNL analysis was then extended to the more complex case of a double--crest wave dynamics and to the resulting single--to--double crest wave transition. The overall agreement with experiments and, especially, the improvements with respect to the simple straightforward asymptotic model are remarkable in all cases considered, although the slight asymmetry observed in the reconstruction of the periodic free surface dynamics at the sidewall was not retrieved in the present model.\\
\indent To the knowledge of the authors, a formal amplitude equation describing the super-harmonic DC sloshing dynamics in orbital shaken containers and coupled with a thorough experimental validation, has not been reported in the literature yet, hence representing the most significant finding of this work.\\
\indent Lastly, by analogy with the close-to-harmonic-resonance dynamics for SC waves, for which the Duffing oscillator represent the suitable mechanical analogy, a one-degree-of-freedom (1dof) mechanical oscillator having both quadratic and cubic nonlinear terms, commonly referred to as Helmholtz-Duffing (HD) oscillator, driven super-harmonically, was tentatively identified as the simplest possible mechanical system that could mimic, at least qualitatively, the super-harmonic DC sloshing dynamics investigated in this paper. \textcolor{black}{The HD equation was largely adopted in the last few decades within the context of structural analysis, i.e. beams, plates and shells subjected to an initial static curvature as well as suspended elastic cables \citep{nayfeh1984quenching,benedettini1989planar}, and it was here proposed as direct mechanical analogy with the present orbital sloshing system.}\\
\indent The main limitation of the models derived in this work is intrinsic to the fundamental assumption of an inviscid fluid. This precludes one to correctly account for the jump-down transition experimentally observed for DC waves at low shaking amplitudes and, therefore, for an accurate estimation of the maximum amplitude response when such a transition occurs. Furthermore, in absence of viscous boundary layers, the weakly nonlinear time-- and azimuthal--averaged mean flow reduces to a free surface deformation only. This is in stark contrast with existence of the so-called Eulerian mean flow \citep{van2018stokes}, also known as viscous streaming flow, typically observed in experiments \citep{bouvard2017mean}. Therefore, the present work overlooks one of the essential points of interest in applications of orbital shaking. The mean flow, which contributes to an efficient mixing, is not captured.\\
\indent The extension of the asymptotic models developed in this work to a viscous analysis is desirable, as it would enable one to predict quantitatively these secondary but fundamental effects for both cases of harmonic and super-harmonic resonances. However, it presently hinges on the subtle modeling of the moving contact line condition.

\appendix

\section{Heuristic damping model: jump--down frequency and DC dynamics suppression at low driving amplitudes}\label{sec:AppA}

\indent In \S\ref{subsubsec:Sec3sub2subsub2} the weakly nonlinear (WNL) model for double--crest (DC) waves was compared with experimental measurements from \cite{reclari2013hydrodynamics} and \cite{reclari2014surface} in terms non-dimensional maximum crest-to-trough contact line amplitude, $\Delta\tilde{\delta}$, for different non-dimensional shaking diameters, $\tilde{d}_s$, and container diameters, $D$ (see figures~\ref{fig:Fig3} and~\ref{fig:Fig31}). We have observed that at larger shaking amplitudes, $\tilde{d}_s$, and for larger container diameter, $D$, a DC wave first emerges on the top of a single--crest (SC) wave at $\Omega\approx\omega_{21}/2$ and eventually wave breaking occurs at larger frequencies. On the contrary, a jump-down transition from DC to SC then takes place by increasing $\Omega$ at lower values of $\tilde{d}_s$ and/or for smaller $D$. 
The latter well-known hysteretic behaviour can be ascribed to the viscous dissipation of the system, obviously overlooked by the present inviscid analysis. In this Appendix, viscous dissipation is tentatively reintroduced by employing a simple heuristic viscous damping model, as described in the following.\\
\indent The viscous dissipation essentially arises at three locations, (i) at the solid tank boundary layers, i.e. bottom and sidewall, (ii) in the fluid bulk and (iii) at the free surface, the latter being typically negligible for ideal surface waves (in absence of any form of contamination). A well-known formula for the prediction of the viscous damping coefficient of capillary--gravity waves in upright cylindrical containers was provided by \cite{Case1957} and \cite{Miles67}. Such an estimation is computed according to the following formula
\begin{equation}
\label{eq:CPdamp}
\sigma=\frac{2k_{mn}}{Re}+\left(\frac{\omega_{mn}}{2Re}\right)^{1/2}\frac{k_{mn}}{\sinh{\left(2k_{mn}H\right)}}+\left(\frac{\omega_{mn}}{2Re}\right)^{1/2}\left[\frac{1}{2}\frac{1+\left(m/k_{mn}\right)}{1-\left(m/k_{mn}\right)}-\frac{k_{mn}H}{\sinh{\left(2k_{mn}H\right)}}\right],
\end{equation}
where the first term represents the bulk dissipation, whereas the second and third terms are related to the dissipation occurring at the solid bottom and sidewall, respectively. In equation~\eqref{eq:CPdamp}, $H=h/R$ is the non-dimensional fluid depth, $k_{mn}$ is the non-dimensional wavenumber associated with mode $\left(m,n\right)$, $\omega_{mn}$ is the corresponding natural frequency obeying to the dispersion relation~\eqref{eq:DispRel} and $Re=g^{1/2}R^{3/2}/\nu$ is the Reynolds number ($\nu$ denotes the kinematic viscosity of the fluid). In \S\ref{subsubsec:Sec3sub2subsub2} an amplitude equation, governing the dynamics of a natural mode $\left(2,n\right)$ (which leads the DC wave dynamics observed close to $\Omega\approx\omega_{21}/2$), was derived. For mode $\left(2,1\right)$ in the conditions of figure~\ref{fig:Fig3}, i.e. pure water with $\rho=1000\text{kg}/\text{m}^3$, $\gamma=0.072\,\text{N}/\text{m}$, $\nu=1\times10^{-6}\,\text{m$^2$}/\text{s}$, $D=0.144\,\text{m}$ (for which the Bond number is $Bo=705.6$) and $H=1.04=2\tilde{H}$, the values $Re=60\,480$, $k_{21}=3.0542$ and $\omega_{21}=1.7561$ give a non-dimensional viscous damping coefficient $\sigma=0.0051$, mostly produced by the sidewall boundary layer. Typically, as in the present case and as supported by experimental \citep{Cocciaro93} and numerical \citep{Viola2018a} evidences, the viscous damping rate can be interpreted as a slow damping process over a faster time scale represented by the wave oscillation. Under this hypothesis, which translates in the assumption of a viscous damping coefficient of order $\epsilon^2$ within the present WNL framework, the damping coefficient can be added \textit{a posteriori}, i.e. in a phenomenological way, to the final inviscid amplitude equation from~\eqref{eq:AmpEqDCfinal}, leading to 
\begin{equation}
\label{eq:AmpEqDCfinal_damp}
\frac{dB}{dt}=-\left[\sigma + \text{i}\left(2\lambda-\chi_{_{DC}} f^2\right)\right] B + \text{i}\left(\zeta_{_{DC}}\lambda+\mu_{_{DC}}\right) f^2 + \text{i}\nu_{_{DC}} |B|^2B.
\end{equation}
The stationary form of~\eqref{eq:AmpEqDCfinal_damp} can be rearranged in the following implicit form
\begin{equation}
\label{eq:AmpEqDCfinal_damp_sol}
\left(2\lambda-\nu_{_{DC}}|B|^2-\chi_{_{DC}}f^2\right)|B|\pm\sqrt{f^4\left(\zeta_{_{DC}}\lambda+\mu_{_{DC}}\right)^2-\left(\sigma|B|\right)^2}=0,
\end{equation}
\begin{figure}
\includegraphics[width=0.5\textwidth]{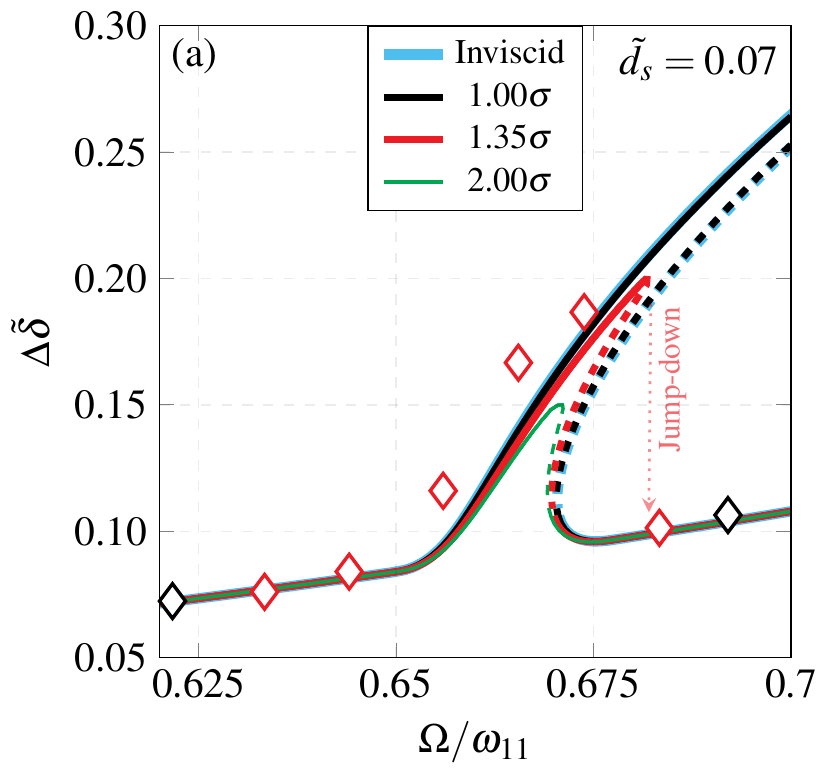}
\includegraphics[width=0.5\textwidth]{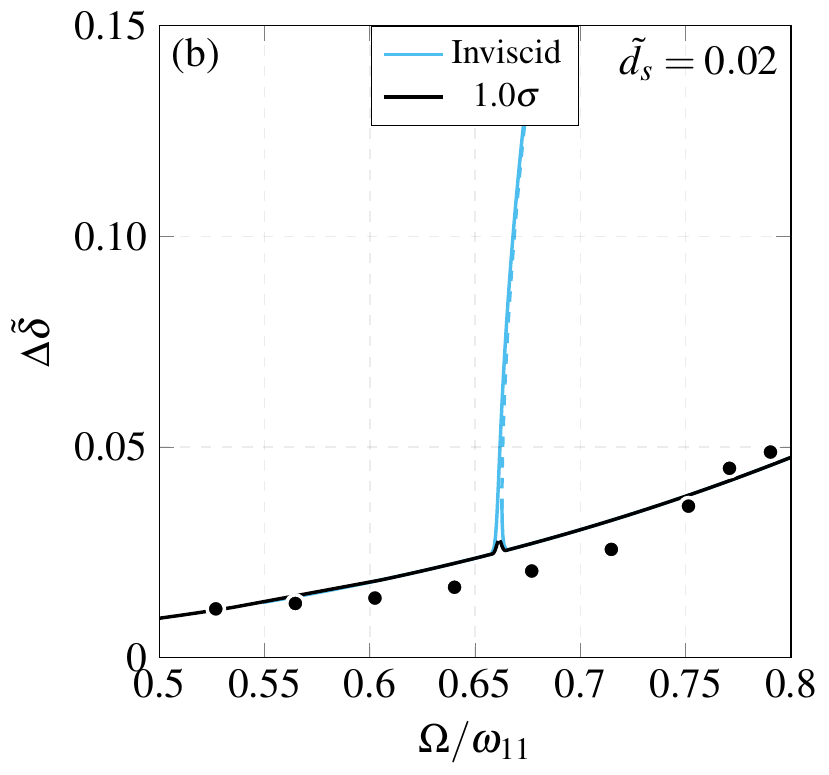}
\caption{(a) Same case of figure~\ref{fig:Fig3}-(a) with $\tilde{d}_s=0.07$. (b) same as (a), but for $\tilde{d}_s=0.02$ (from figure~\ref{fig:Fig1}), value at which the double--crest dynamics does not manifest. Solid and dashed lines correspond to stable and unstable branches, respectively, computed via the weakly nonlinear analysis in the inviscid case and for different values of damping coefficient, with $\sigma$ given by~\eqref{eq:CPdamp}. Markers correspond to the experimental points shown in figure~\ref{fig:Fig1} and extracted from \cite{reclari2014surface}.}
\label{fig:Fig7} 
\end{figure}
which can be solved using the Matlab function \textit{fimplicit}. The effect of viscous dissipation on the DC regime is investigated in figure~\ref{fig:Fig7} for two representative values of the shaking diameter.\\
\indent The case of figure~\ref{fig:Fig7}(a) shows that the so-called jump-down frequency is somewhere in between $\Omega\in\left[0.675,0.685\right]$. The damping value produced by~\eqref{eq:CPdamp} appears to be too small to match the experimental jump-down frequency, hence we tentatively added a pre factor in order to fit the measurements. It turns out that a pre factor of 1.35 is sufficient to provide a fairly good prediction of the jump-down frequency. We note that prediction~\eqref{eq:CPdamp} does not involve any dissipation mechanism associated with the contact line, i.e. contact line hysteresis \citep{Miles67,Cocciaro93,Dussan79,Hocking87,Keulegan59,Kidambi2009,Viola2018a,Viola2018b} %\citep{ngarzone21proj}
or possible surface contamination \citep{henderson1990single,henderson1994surface}. Indeed, depending on the configuration, contact line dynamics may rule the overall dissipation, with a measured damping coefficient up to 10-20 times larger \citep{Benjamin54,Hocking87,Kidambi2009} than that predicted by~\eqref{eq:CPdamp}. Comparison of the theoretical damping coefficient value with that measured in moving contact line experiments, due to unavoidable sources of uncertainty in the meniscus dynamics, have always been mostly qualitative, rather than quantitative, requiring often the use of fitting parameters. For instance, in their predictive theory for single-mode Faraday experiments, \cite{henderson1990single} used an effective fluid viscosity 3 times larger than the actual one. Recently, \cite{bauerlein2021phase} have measured the damping coefficient of the first anti-symmetric sloshing mode in a quasi-two-dimensional rectangular container, which was seen to be approximately 1.5 larger than that predicted by the theory \citep{faltinsen2005liquid}. The need for a pre factor of 1.35 in figure~\ref{fig:Fig7}(a), which approximately corresponds to a fictitious fluid with a dynamic viscosity 1.8 time larger, is therefore not surprising when the damping is computed via~\eqref{eq:CPdamp} and contact line dissipation is neglected.\\
\indent We remark that the reasonings outlined in this Appendix in order to elucidate the effect of viscosity are in fact only qualitative. Many aspects are ignored in the present inviscid analysis with phenomenological damping, two of which are commented in the following.\\
\indent Prediction~\eqref{eq:CPdamp} is only valid for free capillary--gravity waves, whereas dissipation rates of forced wave motions are generally more complex. A proper viscous WNL analysis would produce complex eigenmodes and responses (due to the phase shift owing to viscosity) and hence complex-valued normal form coefficients. For instance, among these coefficients, the imaginary part of $\nu_{_{DC}}$ (or $\nu_{_{SC}}$) multiplied by $|B|^2$ in~\eqref{eq:AmpEqDCfinal_damp}, could be interpreted as a sort of nonlinear damping \citep{Douady90}, $\left(\sigma+\text{Im}\left[\nu\right]|B|^2\right)$, whose contribution to the overall dissipation mechanisms is expected to increase at larger wave amplitudes, hence influencing the location of the jump-down frequency. In contradistinction with the case of a pinned (or fixed) contact line% \citep{bongarzone2021faraday}
, a formal viscous analysis undertaking the case of a moving contact line would require the introduction of a slip length model in order to regularize the well-known contact line stress-singularity \citep{Huh71,Davis1974,Lauga2007,navier1823memoire,Viola2018b}.\\
\indent Most importantly, the inviscid WNL model is not capable to describe the continuous modulation of the phase lag between the external forcing and the wave amplitude response, which has been recently demonstrated by \cite{bauerlein2021phase} (for uni-directional sloshing waves in three-dimensional rectangular container) to be of crucial importance in the correct prediction of the jump-down frequency, otherwise often inaccurate, even when the considered damping coefficient is that measured experimentally. \textcolor{black}{In principle, a formal viscous analysis, as briefly introduced above, is expected to correctly capture such a phase lag.\\
\indent Another interesting case, that is worth to be commented, is that shown in figure~\ref{fig:Fig7}(b). At a shaking diameter $\tilde{d}_s=0.02$ (the lowest reported in figure~\ref{fig:Fig1}), the DC dynamics was not observed at all. This is in conflict with the inviscid straightforward asymptotic analysis, which always prescribes a divergent behaviour close to the dominant super-harmonic, $\Omega\approx\omega_{21}/2$, even for vanishing $\tilde{d}_s$. However, as soon as viscous dissipation is introduced, the energy pumped into the system is not sufficient to overcome dissipative effects and DC waves are essentially suppressed, with a system responses that follows satisfactorily the linear solution (see figure~\ref{fig:Fig1}) showing a single--crest dynamics ranging over the whole frequency window, $\Omega/\omega_{11}\in\left[0,1\right]$, in agreement with experimental evidences.}\\
%\indent As a final side comment, we also note that, in addition to the aforementioned effects, the characteristics of the response in the multi-solution range, i.e. maximum response, jump-up and jump-down frequency, may strongly depend on the frequency sweep rate, which collaborates to determine the energy orbit that the final response will capture by changing the basin of attraction of the high-energy orbit during the sweep. See \cite{park2011slow}, \cite{bourquard2019comment} and \cite{yu2020capture} for a thorough discussion. 

\section{Asymptotic harmonic solution of the undamped Duffing equation}\label{sec:AppB}

\bigskip 

\indent By analogy with the weakly nonlinear analysis for harmonic single--crest wave dynamics presented in \S\ref{subsec:Sec3sub1}, we look for an asymptotic solution of the undamped Duffing equation
\begin{equation}
\label{eq:DuffingEq_AppA}
\ddot{x}+x+ c_3 x^3 =p\cos{\Omega t},
\end{equation}
having the form $x=x_0+\epsilon x_1$. Additionally, as standard in asymptotic solutions of the Duffing equation, we assume a small external forcing amplitude, $p=\epsilon\hat{p}$ and detuning from the exact resonance, i.e. $\Omega=1+\lambda=1+\epsilon\hat{\lambda}$, small nonlinearities through $c_3=\epsilon\hat{c}_3$ and the existence of a characteristic slow time scale $\hat{t}_1=\epsilon t$. Under these assumptions, the $\epsilon^0$--order homogeneous solutions simply reads
\begin{equation}
\label{eq:DuffingEq_eps1_sol}
x_0=C\left(\hat{t}_1\right)e^{\text{i} t}+c.c.\,.
\end{equation}
with $C\left(\hat{t}_1\right)$ to be determined at next order.  At order $\epsilon$ one can readily verify that, in order to avoid secular terms, a solvability condition must be satisfied. Such a condition leads to the very classical amplitude equation
\begin{equation}
\label{eq:DuffingEq_AmpEq}
dD/dt=-\text{i}\,\lambda D + \text{i}\,\left(-1/4\right) p+\text{i}\,\left(3c_3/2\right)|D|^2D,
\end{equation}
where the change of variable $C=De^{\text{i}\lambda t}$ was introduced and each quantity was recast in terms of the corresponding physical value (to eliminate the implicit small parameter $\epsilon$).\\
\indent By noticing that 
\begin{equation}
\label{eq:DuffingEq_AmpEq_Coeff}
-1/4\ \leftrightarrow\ \mu_{_{SC}},\ \ \ \ \ 3c_3/2\ \leftrightarrow\ \nu_{_{SC}},
\end{equation}
one immediately recognizes that equation~\eqref{eq:AmpEqSCfinal} has indeed the same structure of the formal amplitude equation~\eqref{eq:DuffingEq_AmpEq}, thus suggesting that the continuous sloshing system and the one-degree-of-freedom (1dof) Duffing system, under the specific conditions listed above, behave essentially in the same way.\\

\section{Asymptotic super-harmonic solution of the undamped Helmholtz--Duffing equation}\label{sec:AppC}

\bigskip

In this Appendix, although with the additional assumption of vanishing damping, we briefly summarize the super-harmonic weakly nonlinear solution of the Helmholtz--Duffing equation,
\begin{equation}
\label{eq:HD_Eq_AppB}
\ddot{x}+x+c_2 x^2 + c_3 x^3= p\cos{\Omega t},
\end{equation}
\noindent derived by \cite{benedettini1989planar} and introduced in \S\ref{subsec:Sec3sub1}.\\
\indent We look for an asymptotic solution of the form $x=x_0+\epsilon x_1+\epsilon^2 x_2$, to equation~\eqref{eq:HD_Eq_AppB} with $\sigma=0$ (undamped oscillator), $2\Omega=1+\lambda=1+\epsilon\hat{\lambda}$ and with small nonlinearities through $c_2=\epsilon\hat{c}_2$ and $c_3=\epsilon^2\hat{c}_3$ (with the cubic term one order smaller than the quadratic one). The existence of a two slow time scales is hypothesized, $\hat{t}_1=\epsilon t$ and $\hat{t}_2=\epsilon^2 t$. Under these assumptions, the solution of the $\epsilon^0$--order forced linear problem reads
\begin{equation}
\label{eq:HD_Eq_eps1_sol}
x_1=C\left(\hat{t}_1,\hat{t}_2\right)e^{\text{i} t}+f e^{\text{i}\left(1/2\right) t}e^{\text{i}\left(\hat{\lambda}/2\right) \hat{t}_1}+c.c.,
\end{equation}
with $f=\left(2/3\right)p$ and $C\left(\hat{t}_1,\hat{t}_2\right)$ to be determined at next order. At orders $\epsilon$ and $\epsilon^2$, resonating terms produced by the weak quadratic and cubic nonlinearities, respectively, arise, thus requiring the imposition of two solvability conditions prescribing that amplitude $C\left(\hat{t}\right)$ must obey to the following normal forms
\begin{subequations}
\begin{equation}
\label{eq:HD_Eq_AmpEq_int1}
\boxed{\epsilon^1}\text{ :}\ \ \ \ \ \ \ \ \ \ \ \ \ \ \ \ \ \ \ \ \ \ \ \ \ \ \ \ \ \ \ \ \ \ \ \ dC/d\hat{t}_1=\text{i}\,\left(c_2/2\right)f^2e^{\text{i}\hat{\lambda}\hat{t}_1},
\end{equation}
\begin{equation}
\label{eq:HD_Eq_AmpEq_int2}
\boxed{\epsilon^2}\text{ :}\ \ \ \ \ dC/d\hat{t}_2=-\text{i}\,\hat{\lambda}\left(c_2/4\right)e^{\text{i}\hat{\lambda}\hat{t}_1}-\text{i}\,c_5f^2A-\text{i}\,4c_4|C|^2C,
\end{equation}
\end{subequations}
with the full expression of the auxiliary coefficients $c_4$ and $c_5$ (both functions of $c_2$ and $c_3$), given in \cite{benedettini1989planar}. Combining~\eqref{eq:HD_Eq_AmpEq_int1} and~\eqref{eq:HD_Eq_AmpEq_int2} into a single amplitude equation (by summing the two expression by their respective weights, i.e. $\epsilon$ and $\epsilon^2$, and reintroducing the physical quantities in order to eliminate the dependence on the implicit small parameter $\epsilon$), one obtains 
\begin{equation}
\label{eq:HD_Eq_AmpEq_App}
dD/dt=-\text{i}\,\left(\lambda +c_5f^2 \right)D +\text{i}\,\left(1-\lambda/2\right)c_2f^2/2-\text{i}\,4c_4|D|^2D,
\end{equation}
with $C=De^{\text{i}\lambda t}$. Note that the procedure used in the perturbation analysis above and outlined in \cite{benedettini1989planar} is in fact equivalent to that followed in \cite{nayfeh1984quenching} for treating the same second order super-harmonic resonance in a more general case of a two-term excitation. By comparing the various terms of~\eqref{eq:HD_Eq_AmpEq_App} with those of~\eqref{eq:AmpEqDCfinal}, the analogy is evident, thus suggesting that the actual inviscid sloshing dynamics in the double--crest wave regime may be, at least qualitatively, described by the undamped Helmholtz--Duffing equation~\eqref{eq:HD_Eq} driven super-harmonically.

%%%%%%%%%%%%%%%%%%%%%%%%%%%%%%%%%%%
%%%%%%%%%%%%%%%%%%%%%%%%%%%%%%%%%%%
%%%%%%%%%%%%%%%%%%%%%%%%%%%%%%%%%%%

\subsubsection*{\textbf{\textup{Acknowledgements}}}
\indent  \textcolor{black}{We acknowledge Mohamed Farhat for fruitful discussions.}

\subsubsection*{\textbf{\textup{Funding}}}
\indent We acknowledge the Swiss National Science Foundation under grant 200021\_178971.\\

\subsubsection*{\textbf{\textup{Declaration of Interests}}}
\indent The authors report no conflict of interest.

%%%%%%%%%%%%%%%%%%%%%%%%%%%%%%%%%%
%%%%%%%%%%% BIBLIOGRAPHY %%%%%%%%%%%%%%
%%%%%%%%%%%%%%%%%%%%%%%%%%%%%%%%%%

\bibliographystyle{jfm}
 %Note the spaces between the initials
\bibliography{Bibliography}

\end{document}